\documentclass{aims}

\usepackage{amssymb,amsmath,amsthm,mathtools}
\usepackage{graphicx}

\usepackage{hyperref}

\usepackage{xcolor}
\usepackage{bbm}

\mathtoolsset{showonlyrefs}

\newcommand{\numberset}{\mathbb}
\newcommand{\R}{\numberset{R}}

\newcommand{\N}{\numberset{N}}
\newcommand{\Z}{\numberset{Z}}
\newcommand{\dd}{\partial}
\newcommand{\D}{\nabla}
\newcommand{\norm}[1]{\left\lVert#1\right\rVert}

\allowdisplaybreaks

\newcommand{\weakly}{\rightharpoonup}
\newcommand{\weakstar}{\stackrel{\star}{\weakly}}   
\newcommand{\weakstarBV}{\overset{\star}{\rightharpoonup}_{BV}}
\newcommand{\Ltheta}{\overset{{\theta}}{\longrightarrow}}
\newcommand{\LTheta}{\overset{{\Theta}}{\longrightarrow}}

\newcommand{\prodscal}[2]{{#1} \cdot {#2}}
\newcommand{\abs}[1]{\left \vert {#1} \right \vert}

\theoremstyle{plain}
\begingroup
\newtheorem{theorem}{Theorem}[section]
\newtheorem{lemma}[theorem]{Lemma}
\newtheorem{proposition}[theorem]{Proposition}
\newtheorem{corollary}[theorem]{Corollary}
\endgroup

\theoremstyle{definition}
\begingroup
\newtheorem{definition}[theorem]{Definition}
\newtheorem{remark}[theorem]{Remark}

\endgroup

\usepackage{txfonts}
\def\typeofarticle{Research article}
\def\currentvolume{x}
\def\currentissue{x}
\def\currentyear{2023}

\def\ppages{xxx--xxx.}
\def\DOI{...}
\def\Received{...}
\def\Revised{...}
\def\Accepted{...}
\def\Published{...}

\numberwithin{equation}{section}

\makeatletter
\renewcommand{\@biblabel}[1]{#1\hfill \hspace{-0.2cm}}
\makeatother

\begin{document}

\title{Variational Analysis in one and two dimensions of a frustrated spin system: chirality and magnetic anisotropy transitions}

\author{%
  Andrea Kubin\affil{1},
  Lorenzo Lamberti\affil{2}
  }

\shortauthors{the Author(s)}

\address{%
  \addr{\affilnum{1}}{Zentrum Mathematik-M7, Technische Universit\"at M\"unchen, Boltzmannstrasse 3, 85747, Garching, Germany}
  \addr{\affilnum{2}}{Dipartimento di Matematica, Via Giovanni Paolo II 132, 84084, Fisciano, Italy}}

\corraddr{andrea.kubin@tum.de.}

\begin{abstract}
We study the energy of a ferromagnetic/antiferromagnetic frustrated spin system where the spin takes values on two disjoint circles of the 2-dimensional unit sphere. This
			analysis will be carried out first on a one-dimensional lattice and then on a two-dimensional lattice. The energy consists of the sum of a term that depends on nearest and next-to-nearest interactions and a penalizing term related to the spins' magnetic anisotropy transitions. We analyze the asymptotic behaviour of the energy, that is when the system is close to the helimagnet/ferromagnet transition point as the number of particles diverges. In the one-dimensional setting we compute the $\Gamma$-limit of scalings of the energy at first and second order. As a result, it is shown how much energy the system spends for any magnetic anistropy transition and chirality transition. In the two-dimensional setting, by computing the $\Gamma$-limit of a scaling of the energy, we study the geometric rigidity of chirality transitions.
\end{abstract}

\keywords{
$\Gamma$-convergence; frustrated lattice systems; chirality transitions
	}

\maketitle

\section{Introduction}

\begin{figure}
		\centering
		\includegraphics[width=0.3		\textwidth]{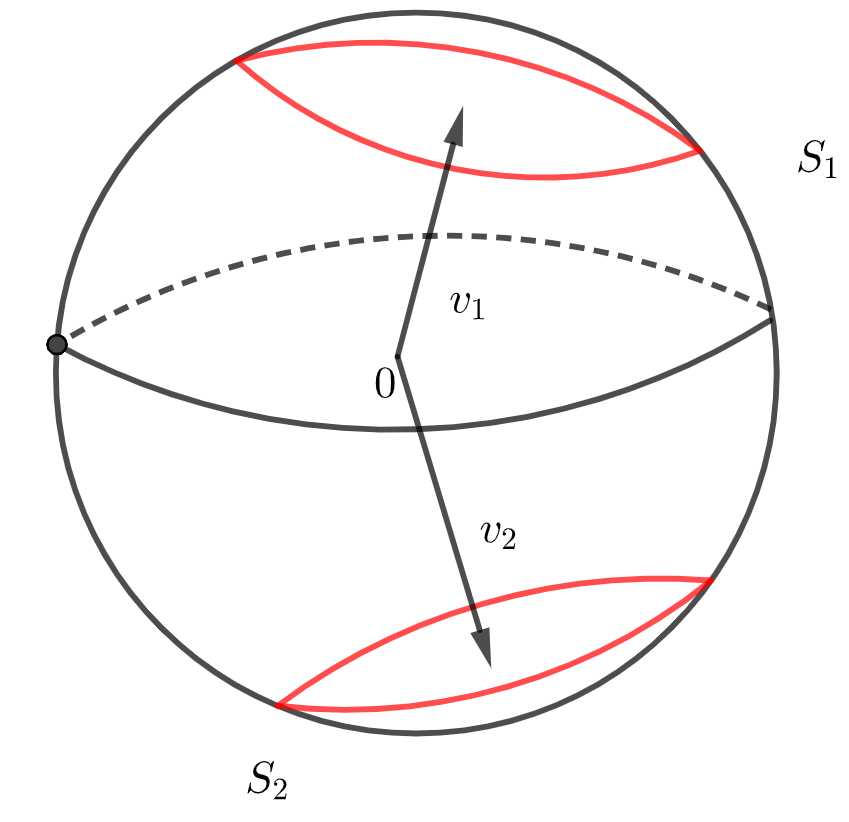}
		\caption{$S_1$ and $S_2$ {circles} of anisotropy transitions.}\label{figura1}
	\end{figure}

Lattice systems are discrete variational models, whose energy depends on a
		spin field defined in a lattice. In frustrated lattice systems, spins cannot find an orientation that simultaneously minimizes the nearest-neighbor (NN) and the next-nearest-neighbor (NNN) interactions. Such interactions are said to be ferromagnetic or antiferromagnetic if they favour alignment or anti-alignment (we address the reader to \cite{diep} for a complete dissertation).\\
	\indent Three-dimensional frustrated magnets generally exist in the magnetic diamond and pyrochlore lattices (see \cite{Dis}) and edge-sharing chains of cuprates provide a natural example of frustrated lattice systems (see \cite{Dre-etal}). Furthermore, jarosites are the prototype for a spin-frustrated magnetic structure, because these materials are composed exclusively of kagomé layers (see \cite{Noc}).\\
		\indent A different frustration mechanism can also be caused by magnetic anisotropy, as it is common in spin ices (see \cite{GMc}). Magnetic anisotropy refers to the dependence of the magnetization of a material on the direction of the applied magnetic field, which acts as a potential barrier (we address the reader to \cite{skomski2008} for a comprehensive overview of magnetism, including a chapter on magnetic anisotropy and the energy barrier). The interplay between the two frustration mechanisms may result in very complicated Hamiltonians (see \cite{Pov}). Most recently, the physics community attempts to find new fundamental effects such as the magnetization plateaus and the magnetization jumps which
		represent a genuine macroscopic quantum effect. For example, kagomé staircases have been of particular interest  because of the concurrent presence of both highly frustrated lattice and strong quantum fluctuations (see \cite{SAAB}).\\
	\indent In this paper we study a frustrated lattice spin system whose spins take values on the unit sphere of $\R^3$. 
	More precisely, a spin of the system $u$ is a vectorial function whose codomain is the union of two fixed disjoint circles, $S_1$ and $S_2$,  of the unit sphere, which have the same radius $R$ and are identified by two versors, $v_1$ and $v_2$, Figure \ref{figura1}. We set the problem in one and two dimensions: in the one-dimensional case (Section \ref{Analysis of the one-dimensional model}) spin fields are parametrized over the points of the discrete set $[0,1]\cap\lambda_n\mathbb{Z}$ and satisfy a periodic boundary condition; in the two-dimensional case (Section \ref{Analysis of the two-dimensional model}) they are parametrized over the points of the discrete set $\Omega\cap\lambda_n\mathbb{Z}^2$, where $\Omega\subset\R^2$ is an open bounded regular domain. In both cases $\{\lambda_n\}_{n\in\N}$ is a vanishing sequence of lattice spacings. In the first setting, the energy of a given spin of the system $u\colon \lambda_n i\in[0,1]\cap\lambda_n\mathbb{Z}\rightarrow u^i\in S_1\cup S_2$ is
	\begin{equation*}
		\mathcal{E}_n(u)=E_n(u)+P_n(u),
	\end{equation*}
	with
	\begin{equation*}
		E_n(u)=\sum_{i\in[0,1]\cap\lambda_n\mathbb{Z}}\lambda_n\left[ -\alpha\prodscal{u^i}{u^{i+1}}+\prodscal{u^i}{u^{i+2}}\right ] \quad\text{and}\quad
		P_n(u)=\lambda_nk_n|D\mathcal{A}(u)|(I),
	\end{equation*}
	where $\alpha\in(0,+\infty)$ is the frustration parameter of the system that  rules the NN and NNN interactions and $\{k_n\}_{n\in\N}$ is a divergent sequence of positive numbers. 
	The term $\mathcal{A}(u)$ indicates the spins' magnetization direction (the so-called magnetic anisotropy) in the two circles. If the number of magnetic anisotropy transitions, i.e. the number of the jumps between the two circles, is finite, $\mathcal{A}(u)$ is a $BV$ function and $|D\mathcal{A}(u)|(I)$ counts them. According to physical considerations, we require that the energy $P_n$ gives a penalizing contribution to the total energy.\\
	\indent It is easy to see that while the first term of the energy $E_n$ is ferromagnetic and favors the alignment of neighboring spins, the second one, being antiferromagnetic, frustrates it as it favors antipodal next-to-nearest neighboring spins. A more refined analysis, contained in Proposition \ref{PropMINIMI} and Remark \ref{Rem}, shows that, for $n$ sufficiently large, the ground states of the system take values on one of the two {circles} and for $\alpha\geq 4$  {are} ferromagnetic (the {spins} {are} made up of {aligned} vectors), while for $0<\alpha\leq 4$ they {are} helimagnetic (the spins consists in rotating vectors with a constant angle ${\phi}=\pm\arccos(\alpha/4))$. The property of the latter case is known in literature as chirality simmetry: the two possible choices for the angle correspond to either clockwise or counterclockwise spin rotations, or in other words to a positive or a negative chirality.

	In this paper, we address a system close to the ferromagnet/helimagnet transition point {(see \cite{DmiKri})}, that is when $\alpha$ is close to 4 {from below}. {We also require that $\lambda_n k_n$ is close to some positive value (that can be also infinite). This assumption is reasonable, since from a physical point of view the change of the spin's polarization involves a larger amount of energy.} Our aim is to provide a careful description of the admissible states and compute their associated energy. In particular, we find the correct scalings to detect the following two {phenomena}: the spins' magnetic anistropy transitions and  chirality transitions that break the rigid simmetry of minimal configurations.
	
	In \cite{CicaleseSolombrino2015}, the authors studied a one-dimensional ferromagnetic/antiferromagnetic frustrated spin {system} with nearest and next-to-nearest interactions close to the helimagnet/ferromagnet transition point as the number of particles diverges. In that case{,} spin fields {take values in the unit circle}. The proposed model is different from that one, where no anisotropy functional $P_n$ was introduced. In {\cite{CicaleseSolombrino2015}} the presence of {a periodic boundary condition} allowed {manipulating $E_n$ in such a way that it can be recast as a discrete version of} a Modica-Mortola type energy, whose $\Gamma$-convergence is well-known in literature (see \cite{modica} and \cite{Mod-Mor}). Indeed, expanding the functional at the first order, under a suitable scaling, {spin fields can} make a chirality transition on a scale of order $\frac{\lambda_n}{\sqrt{\delta_n}}$, when $\frac{\lambda_n}{\sqrt{\delta_n}}$ approaches to a finite nonnegative value, as $n\rightarrow +\infty$ (otherwise no chirality transitions emerge).

	To set up our problem, we let the ferromagnetic interaction parameter $\alpha$ depend on $n$ and be close to 4 from below, that is, we substitute $\alpha$ by $\alpha_n = 4(1-\delta_n)$ for some positive vanishing sequence $\{\delta_n\}_{n\in\N}$. 
	As in \cite{CicaleseSolombrino2015}, the $\Gamma$-limit of $\mathcal{E}_n$ (with respect to the weak$^\star$ convergence in $\mathrm{L}^\infty)$ does not provide a detailed description of the {phenomena} (as a consequence of Theorem \ref{ThmZeroOrder}) and suggests that, in order to get {further} information on the ground states of the system, we need to consider higher order $\Gamma$-limits (see \cite{GCB} and \cite{BraTru}).
	
\begin{figure}
		\centering
		\includegraphics[clip,width=0.6
		\textwidth, height=0.2\textheight]{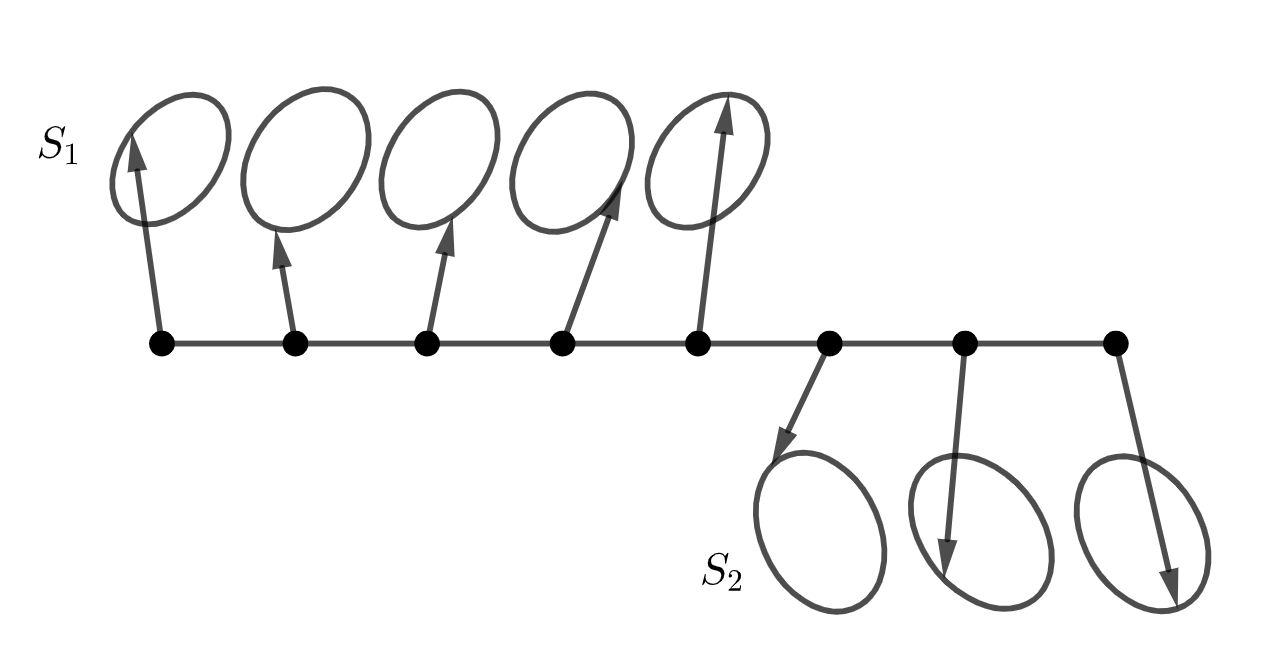}
		\caption{Magnetic anisotropy transitions.}\label{figura2}
	\end{figure}

	\indent The two {phenomena} can be detected at different orders. At the first order we are led to normalize the energy $E_n$ of the system and study the asymptotic behavior of {(a rescaling of)} the new functional ${G}_n$ defined {by}
	\begin{equation*}
		{G}_n={E_n-\min \mathcal{E}_{n}}.
	\end{equation*}
	Rescaling ${G}_n$ by $\lambda_n$, {we prove that} magnetic anisotropy transitions can be {captured when $\lambda_n k_n$ is close to any positive finite value}, for $n$ large enough (see Theorem \ref{Teorema Gamma-convergenza ordine 1}). {At the scale value $\lambda_n$, the energy spent for spin's magnetic anisotropy transitions is equal to the minimal energetic value corresponding to the sum of all the interactions in proximity of the transition points.} In {Figure \ref{figura2} it can be seen an occurrence of the phenomenon that we are analyzing.}\\
	\indent {Chirality transitions can be detected at the next order by means of a technical decomposition of the energy $G_n$. The idea behind the construction in Subsection \ref{Costruzione spin modificati} is to split the problem set in the sphere into {finitely many} problems set in one of the two circles each. We associate each spin field $u$ with a unique and finite partition of $[0,1]$ containing intervals $I_j$ such that $u_{|I_j}$ takes values only in one circle. We note that the intervals $I_j$ depend on $n$ because $u$ is defined on the lattice $[0,1]\cap\lambda_n\mathbb\Z$.} 
	We modify such restrictions $u_{|I_j}$ in {such} a way that they {still satisfy a similar periodic boundary condition on $I_j$}, {denoting} them as $\widetilde{u}_{I_j}$. 
	{In Lemma \ref{DecompGn} we decompose the functional ${G}_n$ as follows:} 
	\begin{equation*}
		{G}_n(u)= \sum_{j} MM_n(  \widetilde{u}_{I_j})+ \sum_{j} (R_n)_j(u) {+(R_n)_{M(u)}(u)+R_n(u).}
	\end{equation*}
	
	{The energy $MM_n$ is of discrete Modica-Mortola type and collects the pairwise interactions of spins' vectors pointing to the same circle; the functionals $(R_n)_j$ and $(R_n)_{M(u)}$ gather the interactions between consecutive spins' vectors that point to different circles. $R_n$ is a correction addend}. The {first sum and the other addend in the right-hand side of the previous formula} need to be rescaled in different ways, the first sum {being} a higher order term. Thus, at the second order we deal with the energy
	\begin{equation*}
		{\mathcal{G}}_n(u)={G}_n(u) -\sum_{j} (R_n)_j(u){-(R_n)_{M(u)}(u)-R_n(u)}= \sum_{j} MM_n(  \widetilde{u}_{I_j} ).
	\end{equation*}
	{In Theorem \ref{Teoremafinale} we apply the $\Gamma$-convergence result contained in \cite{CicaleseSolombrino2015} to each functional $MM_n$, rescaled by $\lambda_n\delta_n^{3/2}$. It turns out that different scenarios may occur, depending on the value of $\lim_n \lambda_n/\sqrt{\delta_n}:= l \in [0,+\infty]$}. If $l =+\infty$, chirality transitions are forbidden. {Otherwise a spin field can make a chirality transition on a lenght-scale $\lambda_n/\sqrt{\delta_n}$. In particular,} if $l>0$, {it} may have diffuse and regular macroscopic (on an order one scale) chirality transitions in each $S_j$ whose limit energy is finite on $H^{1}(I_j)$ (provided some boundary conditions are taken into account){;} if $l=0$, chirality transitions on a mesoscopic scale are allowed. In this case, the continuum limit energy is finite on $BV(I_j)$ and counts the number of jumps of the chirality of the {spin field}.
	
	\indent {Systems defined in planar structures are much more difficult to study, due to the higher dimensional setting (see \cite{ACXY}, \cite{BCKO},  \cite{BadCicDLPon},  \cite{CicOrlRuf}, \cite{CRS}). We address here the two-dimensional analogue of the frustrated spin chain studied in the first part of the paper. The energy of a given spin of the system $u\colon {(i,j)}\in\Omega\cap\lambda_n\mathbb{Z}^2\rightarrow {u^{i,j}\in} S_1\cup S_2$ is
		\begin{equation*}
			\mathcal{E}_n(u;\Omega)=E_n(u;\Omega)+P_n(u;\Omega),
		\end{equation*}
		where
		\begin{equation*}
			E_n(u;\Omega)=-\alpha\sum_{(i,j)}\lambda_n^2(\prodscal{u^{i,j}}{u^{i+1,j}}+\prodscal{u^{i,j}}{u^{i,j+1}})+\sum_{(i,j)}\lambda_n^2(\prodscal{u^{i,j}}{u^{i+2,j}}+\prodscal{u^{i,j}}{u^{i,j+2}})
		\end{equation*}
		and
		\begin{equation*}
			P_n(u;\Omega):=\lambda_nk_n|D\mathcal{A}(u)|(\Omega).
		\end{equation*}
		We assume that the functional $P_n(\cdot;\Omega)$	is bounded. The number $\alpha>0$ is the frustration parameter of the system and $\{k_n\}_{n\in\N}$ is a divergent sequence of positive numbers. The {term} $|D\mathcal{A}(u)|(\Omega)$ is related to magnetic anistropy transitions. In the two-dimensional setting, they occur on the edges of the lattice $\Omega\cap\lambda_n\mathbb{Z}^2$ and the natural number $\frac{|D\mathcal{A}(u)|(\Omega)}{\lambda_n|v_1-v_2|}$ is an upper bound on the spins' transitions from a circle to the other in $\Omega$.\\
		\indent Motivated by the variational analysis of the one-dimensional problem, we assume that the frustation parameter depend on $n$ and is close to the helimagnet/ferromagnet transition point as the number of particles diverges, i.e. $\alpha_n\rightarrow 4^-$. In view of detecting spins' chirality transitions, which cannot be captured by means of the $\Gamma$-limit of the energy at the zero order, we are interested in the functional
		defined by}
	
	\begin{equation*}
		H_n(u;\Omega):=\frac{1}{\sqrt{2}\lambda_n\delta_n^{\frac{3}{2}}}\frac{1}{2}\lambda_n^2\sum_{{(i,j)}}\left[\left| u^{i+2,j}-\frac{\alpha_n}{2}u^{i+1,j}+u^{i,j}\right|^2+\left| u^{i,j+2}-\frac{\alpha_n}{2}u^{i,j+1}+u^{i,j}\right|^2\right],
	\end{equation*}
	{which is the two-dimensional analogue of $G_n$, up to additive constants.}\\
	\indent {In \cite{CicOrlRuf} the authors studied a similar frustrated spin chain whose spin fields take values in the unit circle of $\R^2$. In \cite[Theorem 2.1]{CicOrlRuf} they  proved the emergence of spins' chirality transitions by means of the $\Gamma$-convergence of the functional $H_n$ with respect to the local $\mathrm{L}^1$-convergence of two chirality parameters.\\
		\indent In view of applying their result in our setting, we employ an idea that recalls the construction carried out in the one-dimensional problem. We} restrict every spin $u$ to connected open sets $C_s$ that partition $\Omega$ in such a way that {$u_{|C_s}$} takes values only in one {circle}. In order to avoid more complicated notation, we do not impose boundary conditions on $\partial\Omega$ and we state the result by means of a local convergence. We note that the sets $C_s$ depend on $n$ because $u$ is defined on the lattice $\Omega\cap\lambda_n\mathbb\Z^2$.\\
	\indent {We decompose
		\begin{equation*}	H_n(u;\Omega)=\sum_{s}\Big[\mathcal{H}_n(u;C_s)+(R_n)_{C_s}(u)\Big],
		\end{equation*} 
		where $\mathcal{H}_n$ collects the interactions of spins' vectors {pointing} to the same circle and $(R_n)_{C_s}$ gathers the interactions between spins' vectors that {point to} different circles.}

	\indent
	\indent While in the one-dimensional setting the partition associated with a spin {contains} intervals, which guaranty the compactness results stated, in this case the sets $C_s$ could be very wild, as the spacing of the lattice shrinks. Therefore, we require as additional regularity {condition} for the components $C_s$, that is the $BVG$ regularity. {Its definition can be found in \cite{Pol07} and is recalled in Definition \ref{BVGReg}.}\\
	\indent With this regularity assumption, we can apply the $\Gamma$-convergence result proved in \cite{CicaleseFosterOrlando2019} {to each addend of} the functional
	{
		\begin{align*}
			G_n(h;\Omega)=
			H_n(h,\Omega)-\sum_s (R_n)_{C_s}(h)=\sum_s\mathcal{H}_n(h;C_s),
		\end{align*}
		as it is shown in Theorem \ref{Teorema principale 2d}, that is the main result of Section \ref{Analysis of the two-dimensional model}. It turns out that chirality transitions are possible and they can take place both in the vertical and horizontal slices of $C_s$. 
	}

	{\section{Basic notation}}
	Given $ x \in \R$, we denote by $ \lfloor x \rfloor $ the integer part of $x$. For a set $K$ we denote {by} $\mathrm{co}(K)$ the convex hull of $K$, {by $\# K$ the number of its elements and $\chi_K$ its characteristic function}.  We {write} $v\cdot w$ {for} the {Euclidean} scalar product of the vectors $v,w\in\R^3$ and by $S^2$ the unit sphere of $\R^3$. For all $ v \in \R^3$ we {denote by $ \pi_v $} the {Euclidean} projection on $ v $ and {by $ \pi_{v^\bot}$} the projection on the orthogonal complement of $v$. {If $A$ is a subset of the Euclidean space we denote by $\overline{A}$ its closure respect the Euclidean topology.}  {We denote by $C$ a generic constant that may vary from line to line in the same formula and between formulas. Relevant dependencies on parameters and special constants
		will be suitably emphasized using parentheses or subscripts.}\\
	\indent  If $I\subset\R$ is an interval and all $w \in BV(I;\R^3)$, we denote {by} $Dw\in \mathcal{M}_b(I; \R^3) $ the distributional differential of $w$, and {by} $ \vert Dw \vert \in \mathcal{M}_b(I)$ the total variation measure of $Dw$. We say that a sequence $\{u_n\}_{n \in \N}$ {converges weakly$^\star$} in $BV(I;\R^3)$ to a function $u \in BV(I;\R^3)$ if and only if
	\begin{equation*}
		u_n \rightarrow u \quad \text{in $\mathrm{L}^1(I;\R^3)$ $ \quad$ and } \quad {\sup_{n\in\N}|Du_n|(I)<+\infty,}
	\end{equation*}
	{(see \cite[Definition 3.11 and Proposition 3.13]{AmbFusPal})}. {We} denote it by $ u_n {\weakstarBV} u$.\\
	\indent Fixing $v_1,v_2\in S^2$ and $ R \in (0,1)$, we define the set
	\begin{equation}\label{DEFS_v}
		S_{i}:=\left\{w\in S^2\,:\, \lvert\pi_{v^\bot_i}(w)\rvert=R, \langle w,v_i	\rangle>0 \right\},\quad\text{for }i\in\{1,2\}.
	\end{equation}
	It is easy to observe that the set $S_i$ is a {circle} centered in $ v_i \sqrt{1-R^2}$ and {it} can be easily verified that for $R<R_{Max}:=\sqrt{\frac{1-\prodscal{v_1}{v_2}}{2}}$ the sets $S_1$ and $S_2$ are disjoint. Throughout the paper we assume that $R \in (0,R_{Max})$.\\
	\indent  {If $S$ is an open set of $\R^N$ and $\mathcal{C}$ is a collection of open subsets of $S$, we say that $\mathcal{C}$ is an open partition of $S$} if $\mathcal{C}$ does not contain empty sets and
	\begin{equation*}
		\overline{S}=\bigcup_{C\in\mathcal{C}}\overline{C},\quad C_1\cap C_2=\emptyset, \quad \forall C_1,C_2\in\mathcal{C}.
	\end{equation*}
	\indent{Given two vectors} $w=(w^1,w^2),\, \overline{w}=(\overline{w}^1,\overline{w}^2)$ of $\R^2$, we define the function
	\begin{equation*}
		\chi[w,\overline{w}]:=\mathrm{sign}(w^1\overline{w}^2-w^2\overline{w}^1),
	\end{equation*}
	{with the convention that $\mathrm{sign}(0)=-1$.}
	$\\ $

	\section{Analysis of the one-dimensional model}
	\label{Analysis of the one-dimensional model}
	\subsection{Further notation and definitions}
	\indent We {let} $I = (0, 1)$ and we consider a sequence $\{\lambda_n\}_{n \in \mathbb{N}}\subset\R^+$ that vanishes as $ n\rightarrow + \infty$. It represents a sequence of spacings of the lattice {$\overline{I}\cap\lambda_n\mathbb{Z}$}.\\
	\indent We introduce the class of functions valued in $S_1\cup S_2$ which are {piecewise constant} on the edges of the lattice {$\overline{I}\cap\lambda_n\mathbb{Z}$ and} satisfy a {periodic} boundary condition:
	\begin{align}\label{funzconttrat13032023}
		\mathcal{PC}_{\lambda_n}:=
		\Big\{ {u}\colon {\overline{I}}\rightarrow S_1\cup S_2\,:\, &{u}(t)= {u}(\lambda_n i) \text{ for }t\in \lambda_n[i+[0,1)]{\cap\overline{I}\text{ and } \lambda_n i\in\overline{I}\cap\lambda_n\mathbb Z,}\notag\\
		&\prodscal{{u}^0}{{u}^1}= \prodscal{ {u}^{ \left \lfloor\frac{1}{\lambda_n} \right \rfloor-1 }  }{{u}^{\left \lfloor\frac{1}{\lambda_n} \right \rfloor} }\Big\}.
	\end{align}
	We will identify a piecewise {constant} function ${u}\colon {\overline{I}}\rightarrow S_1\cup S_2$ with the function defined on the points of the lattice given by ${\lambda_n i\in\overline{I}\cap\lambda_n\mathbb{Z}}\mapsto {u}^i:={u}(\lambda_n i)$. Conversely, given values ${u}^i\in S_1\cup S_2$ for ${\lambda_n i\in\overline{I}\cap\lambda_n\mathbb{Z}}$, we define ${u}\colon {\overline{I}}\rightarrow S_1\cup S_2$ by ${u}(t):={u}^i$ for $t\in \lambda_n[i+[0,1)]$.\\
	\indent There exists a natural projection map $\mathcal{A}\colon\mathrm{L}^\infty(I;S_1\cup S_2)\rightarrow  {\mathrm{L}^\infty(I;\{ v_1,v_2\})}$ defined as follows:
	\begin{equation}\label{functionAcorsivoinie}
		\mathcal{A}(u){(t)}=
		\begin{cases}
			v_1\quad&\text{if }u(t)\in S_1,\\
			v_2 &{\text{if }u(t)\in S_2},
		\end{cases}
		\quad{\forall t\in I.}
	\end{equation}
	{For each spin $u$, the function $\mathcal{A}(u)$ indicates the spins' magnetization direction and its jumps correspond the the spins' magnetic anisotropy transitions.} 
	{In general, $\mathcal{A}$ can be defined analogously on $\mathrm{L}^\infty(I;K_1\cup K_2)$, if $K_1$ and $K_2$ are two disjoint subsets of $\R^3$ containing, respectively, $S_1$ and $S_2$. In this case, we remark that if a spin field $u\in\mathrm{L}^\infty(I;K_1\cup K_2)$ switches from $K_1$ to $K_2$ a finite number of times, i.e. $\mathcal{A}(u)\in BV(I;\{v_1,v_2\})$ and so $|D\mathcal{A}(u)|(I)<+\infty$,} the interval ${I}$ can be partitioned in {finitely many} regions where the function $u$ takes {values} only in one of the two {sets $K_1$ and $K_2$}. In other words, there exist $M
	(u)\in\N$ and a collection of open intervals, $\{{I_j}\}_{j\in\{1,\dots,M(u)\}}$, such that
	\begin{equation}
		\label{Partizionehp11d}
		\{{I_j}\}_{j\in\{1,\dots,M(u)\}}\text{ is an open partition of } I,
	\end{equation}
	\begin{equation}
		\label{Partizionehp21d}
		{\text{either } u(I_j)\subset K_1\text{ or } u(I_j)\subset K_2 ,\text{ for any $j\in\{1,\dots,M(u)$\}}},
	\end{equation}
	\begin{equation}
		\label{Partizionehp31d}
		{u(I_j)\times u(I_{j+1})\subset (K_1\times K_2)\cup (K_2 \times  K_1), \text{for any $j\in\{1,\dots,M(u)-1$\}}.}
	\end{equation}
	{The last two} properties imply that this partition is unique. We observe that{, if $u\in\mathrm{L}^\infty(I;S_1\cup S_2)$ and $\mathcal{A}(u)\in BV(I;\{v_1,v_2\})$ (or, in particular, if $u\in\mathcal{PC}_{\lambda_n}$), then}
	\begin{equation*}
		M(u)=\frac{|D\mathcal{A}(u)|(I)}{|v_1-v_2|}+1.
	\end{equation*}
	The following definition will be useful throughout the section.
	\begin{definition}\label{25052022defC_n}
		{Let $u\in\mathrm{L}^\infty(I;S_1\cup S_2)$ be such that $\mathcal{A}(u)\in BV(I;\{v_1,v_2\})$}. We say that $\mathcal{C}_n(u)=\{ I_j\,|\,j\in\{1,\dots,M(u)\} \}$ is the open partition associated with $u$ if $M(u)=\frac{|D\mathcal{A}(u)|(I)}{|v_1-v_2|}+1$ and the collection of open intervals $\{I_{j}\}_{j\in\{1,\dots,M(u)\}}$ satisfies \eqref{Partizionehp11d}, \eqref{Partizionehp21d} and \eqref{Partizionehp31d}.
	\end{definition}

	\subsection{Some properties of $ \mathrm{L}^\infty $ functions with values in a compact set} 
	\label{Some properties}
	In this subsection we recall some classical properties of the Lebesgue space $ \mathrm{L}^{\infty}(I;K)$, where $K \subset \mathbb{R}^{{3}}$ is a compact set.{ The statements and the proofs are fully analogous if the setting is a $N$-dimensional Euclidean space.}
	\begin{proposition}\label{propLinfliminconv}
		Let $ \{ f_n\}_{n \in \mathbb{N}} \subset \mathrm{L}^{\infty}(I; K)$. Then, up to subsequences, $f_n \weakstar f\in \mathrm{L}^{\infty}(I; \mathrm{co}(K))$ in the {weak$^\star$} topology of $\mathrm{L}^{\infty}(I; \mathbb{R}^{{3}})$. Moreover, for all $u \in \mathrm{L}^{\infty}(I;\mathrm{co}(K))$ there exists a sequence $\{u_n\}_{n \in \N} \subset \mathrm{L}^{\infty}(I;K)$ of piecewise constant functions such that $u_n \weakstar u$. 
	\end{proposition}
	\begin{proof}
		Since the set $K $ is bounded then, up to a subsequence, there exists $f \in \mathrm{L}^{\infty}(I;\mathbb{R}^{{3}})$ such that $f_n \weakstar f$. {Now we} prove that $f(t) \in \mathrm{co} (K)$ for almost every $t \in {I}$. For every $\xi \notin \mathrm{co}(K) $ there exist an affine function $h_\xi : \mathbb{R}^N \rightarrow \mathbb{R}$  and $\alpha<0$ such that 
		\begin{equation*}
			h_\xi(\xi)>0>\alpha>h_\xi(x), \quad \forall x \in \mathrm{co}(K).
		\end{equation*}
		By the {weak$^\star$} convergence of $\{f_n\}_{n \in \mathbb{N}}$ we have that for any measurable set $A \subset {I}$
		\begin{equation*}
			\int_{A} h_\xi(f(t))dt= \lim_{n \rightarrow + \infty} \int_{A} h_\xi(f_n(t))dt\leq  \vert A\vert \alpha<0.
		\end{equation*}
		Hence, by the arbitrariness of $A $, we obtain
		\begin{equation}\label{asdasd1}
			h_\xi(f(t))<0, \quad \text{ for a.e. $t \in {I}$.}
		\end{equation}
		Recalling that 
		\begin{equation*}
			\mathrm{co}(K)= \bigcap_{j \in \mathbb{N}} \left\{ y \in \mathbb{R}^3:\; h_{\xi_j}(y)<0, \; \xi_j \in \mathbb{Q}^N \setminus \mathrm{co}(K) \right\},
		\end{equation*}
		by formula \eqref{asdasd1} we obtain
		\begin{equation*}
			f(t) \in \mathrm{co}(K), \quad \text{ for a.e. $t \in {I}$.}
		\end{equation*}
		
		\indent {Now we prove the second statement of the proposition. 
			Let $u \in \mathrm{L}^{\infty}(I;\mathrm{co}(K))$. There exists a sequence $\{u_n\}_{n \in \mathbb{N}} \subset \mathrm{L}^{\infty}(I;\mathrm{co}(K)) $ such that $u_n= \sum_{j=1}^{m} a_j \chi_{I_j}$, where $a_j \in \mathrm{co}(K)$ and $I_j \subset I$ is an interval, for any $j\in\{1,\dots,m\}$, and $u_n$ converges to $u$ in $\mathrm{L}^1(I;\R^3)$. Hence, $u_n\weakstar u$. Therefore, without loss of generality, we may prove the statement for $u=a\in \mathrm{co}(K)$}.\\
		\indent  We define the following function:
		\begin{equation*}
			h(t):=
			\begin{cases}
				a_1 & \quad \text{if $t \in (0,\lambda)$,}\\
				a_2 & \quad \text{if $ t \in [\lambda,1)$},
			\end{cases}
		\end{equation*}
		where $a= \lambda a_1+ (1-\lambda)a_2$ with $ a_1, a_2 \in K$ and for some $ \lambda \in [0,1]$. Then the sequence $ u_n(t):= h(nt)$ converges to $ u$ in the weak${^\star}$ topology of $ \mathrm{L}^{\infty}$ by Riemann-Lebesgue's lemma.
	\end{proof}
	\begin{corollary}
		The closure of the set $\mathrm{L}^{\infty
		}(I;K)$ with respect to the weak$^\star$ topology of $\mathrm{L}^{\infty}(I;\R^{{3}})$ is the set $ \mathrm{L}^{\infty}(I;\mathrm{co}(K))$.
	\end{corollary}
	\begin{proof}
		Since the space $\mathrm{L}^1(I;\R^{{3}})$ is separable, every bounded subset of $ \mathrm{L}^{\infty}(I;\R^{{3}})$ is metrizable with respect to the weak$^\star$ topology of $ \mathrm{L}^{\infty}(I;\R^{{3}})$. Hence the set $ \mathrm{L}^{\infty}(I;K)$ is metrizable. Therefore, by {Proposition \ref{propLinfliminconv}}, we have that the set $\mathrm{L}^{\infty}(I;\mathrm{co}(K))$ is the weak$^\star$ closure of the set $\mathrm{L}^{\infty}(I;K)$.
	\end{proof}
	
	\subsection{A useful abstract result} In this subsection we cite an abstract $\Gamma$-convergence result proved in \cite{AlicandroCicalseGloria2008} that will be applied in Subsection \ref{GammaConvZero}. For this purpose, we introduce the following notation. Let $K \subset \mathbb{R}^N$ be a compact set and for all $ \xi \in \mathbb{Z}$ let $f^{\xi} \colon \mathbb{R}^{2N} \rightarrow \R$ be a function such that 
	\begin{itemize}
		\item[(\textbf{H1})] $ f^{\xi}(x,y)= f^{-\xi}(y,x)$,
		\item[(\textbf{H2})] for all $ \xi \in \mathbb{Z}$, $ f^{\xi}(x,y)=+\infty$ if $ (x,y) \notin K^2$, 
		\item[(\textbf{H3})] for all $\xi \in \mathbb{Z}$ there exists $C^\xi \geq 0$ such that 
		\begin{equation*}
			\sup_{(x,y) \in K^2} \vert f^{\xi}(x,y) \vert \leq C^{\xi} \quad \text{and} \quad \sum_{\xi \in \mathbb{Z}} C^{\xi} < +\infty.
		\end{equation*}
		
	\end{itemize}
	For any $n \in \N$ we define the functional space
	\begin{equation*}
		D_n(I; \mathbb{R}^{N}) \colon = \left \{  u\colon \mathbb{R} \rightarrow \mathbb{R}^N \, \colon\, u \text{ is constant in } \lambda_n(i+ [0,1)) \text{ for all $ {\lambda_n}i \in {\overline{I}\cap\lambda_n\mathbb{Z}}$}\right\}.
	\end{equation*}
	{With the notation already used, we denote the value of the function $u$ in the interval $ \lambda_n(i+ [0,1))$ by $u^{i}$ for all $\lambda_n i \in \overline{I} \cap \lambda_n \mathbb{Z}$. We introduce} the sequence of {functionals} $F_{n} \colon \mathrm{L}^{\infty}(I;\mathbb{R}^N) \rightarrow (-\infty,+\infty]$ defined as follows:
	\begin{equation}\label{energieNfunct}
		F_{n}(u):=
		\begin{cases}
			\displaystyle\sum_{\xi \in \mathbb{Z}} \sum_{i \in R_{n}^{\xi}(I)} \lambda_n f^{\xi}(u^i,u^{i+\xi}) & \text{ for } u \in D_n(I;\mathbb{R}^N), \\
			+\infty & \text{ for } u \in \mathrm{L}^{\infty}(I;\mathbb{R}^N) \setminus D_n(I;\mathbb{R}^N),
		\end{cases}
	\end{equation}
	where $ R_n^{\xi}(I):= \{ {\lambda_n} i \in {\overline{I}\cap\lambda_n\mathbb{Z}}\, \colon \, {\lambda_n}i + \xi \in {\overline{I}\cap\lambda_n\mathbb{Z}}\}$, {for $\xi\in\mathbb Z$.} For any open and bounded set $A \subset \R$ and for every $ {u} \colon\mathbb{Z} \rightarrow \mathbb{R}^N$, we define the discrete average of ${u}$ in $A$ as
	\begin{equation*}
		({u})_{1,A}:= \frac{1}{\# (\mathbb{Z} \cap A)} \sum_{i\in \mathbb{Z} \cap A} {u}^i.
	\end{equation*}
	
	\begin{theorem}[See \cite{AlicandroCicalseGloria2008}]\label{THMaliCicGlor}
		Let $\{f^{\xi}\}_{\xi \in \mathbb{Z}}$ {be} a family of functions that satisfies \textbf{H1}, \textbf{H2}, \textbf{H3}. Then the sequence $ F_n$ {$\Gamma$}-converges, as $n \rightarrow + \infty$, with respect to the {weak$^\star$} topology of $\mathrm{L}^{\infty}(I;\mathbb{R}^N)$, to 
		\begin{equation*}
			F(u):=
			\begin{cases}
				\displaystyle\int_{I}f_{hom}(u(t))\,dt & \text{ for } u \in \mathrm{L}^{\infty}(I;\mathrm{co}(K)), \\
				+\infty & \text{ for } u \in \mathrm{L}^{\infty}(I;\mathbb{R}^N) \setminus \mathrm{L}^{\infty}(I;\mathrm{co}(K)),
			\end{cases}
		\end{equation*}
		where $f_{hom}\colon \mathbb{R}^N \rightarrow \mathbb{R}$ is given by the following homogenization formula
		\begin{equation*}
			f_{hom}(z)= \lim_{\rho \rightarrow 0} \lim_{k \rightarrow + \infty} \frac{1}{k}  \inf \left\{  \sum_{\xi \in \mathbb{Z}} \sum_{\beta \in R_1^{\xi}((0,k))} f^{\xi}({u}(\beta), {u}(\beta+\xi))\, {\bigg|\, {u} \text{ s.t. }}({u})_{1,(0,k)}  \in \overline{B(z,\rho)} \right\}.
		\end{equation*} 
	\end{theorem}

	\subsection{The energy model and its ground states}
	{This subsection is devoted to the mathematical formulation of the model and the characterization of its ground states.}\\
	\indent Let $\alpha>0$ be a fixed parameter and let $\{k_n\}_{n \in \N}{\subset}\R^+$ be a divergent sequence of positive numbers. Denoting
	\begin{equation}
		\label{InsiemeIndici}\mathcal{I}^n(I):={(\overline{I}\cap\lambda_n\mathbb Z)\setminus\left\{ \lambda_n \left(\left\lfloor\frac{1}{\lambda_n}\right\rfloor-1\right), \lambda_n\left\lfloor\frac{1}{\lambda_n}\right\rfloor\right\},}
	\end{equation}
	{and in general if $J=(a,b) \subset I$ we define 
		\begin{equation} \label{InsiemeIndici14032023}
			\mathcal{I}^n(J):=(\overline{J}\cap\lambda_n\mathbb Z)\setminus\left\{ \lambda_n \left(\left\lfloor\frac{b}{\lambda_n}\right\rfloor-1\right), \lambda_n\left\lfloor\frac{b}{\lambda_n}\right\rfloor\right\}.
	\end{equation}}
	We define the energy of the system as the sum of two addends. The first addend is a bulk scaled energy of a frustrated F-AF spin chain, $E_n \colon\mathcal{PC}_{\lambda_n}\rightarrow(-\infty,+\infty)$, having the following form:
	\begin{equation}\label{energiavecchia}
		E_{n}(u):= \lambda_n\sum_{i\in\mathcal{I}^n(I)}\left[ -\alpha\prodscal{u^i}{u^{i+1}}+\prodscal{u^i}{u^{i+2}} \right].
	\end{equation}
	The second addend of the energy, $P_n \colon \mathcal{PC}_{\lambda_n}\rightarrow[0,+\infty)$, is a term of confinement in $S_1\cup S_2$ and is defined as follows:
	\begin{equation*}\label{cofinamentoEnergia}
		P_{n}(u):= \lambda_n k_n  \left\vert D\mathcal{A}(u) \right \vert(I),
	\end{equation*}
	where $\mathcal{A}$ is the function defined in formula \eqref{functionAcorsivoinie}. We consider the family of energies $\mathcal{E}_n\colon\mathcal{PC}_{\lambda_n}\rightarrow(-\infty,+\infty)$  defined by
	\begin{equation*}
		\mathcal{E}_n(u)=E_n(u)+P_n(u).
	\end{equation*}
	Furthermore, we define the functional $H_n\colon \mathcal{PC}_{\lambda_n} \rightarrow [0,+\infty)$ by
	\begin{align*}
		H_n(u):= \displaystyle \frac{1}{2} \lambda_n \sum_{i\in\mathcal{I}^n(I)} \left \vert u^{i+2}- \frac{\alpha}{2}u^{i+1}+u^{i} \right \vert^{2}. 
	\end{align*}
	If $u\in\mathcal{PC}_{\lambda_n}$, since $ \vert u^i \vert=1$ for all $ {\lambda_n i \in \overline{I}\cap\lambda_n\mathbb{Z}}$, thanks to the boundary condition contained in the definition of $\mathcal{PC}_{\lambda_n}$ {(see \eqref{funzconttrat13032023})}, {we compute:}
	\begin{align}\label{EspressioneHn}
		H_n(u)
		& =\frac{1}{2}\lambda_n\sum_{i\in\mathcal{I}^n(I)}\bigg(2+\frac{\alpha^2}{4}+2\prodscal{u^i}{u^{i+2}}-\alpha\prodscal{u^i}{u^{i+1}}-\alpha\prodscal{u^{i+1}}{u^{i+2}}\bigg)\notag\\
		& = \frac{1}{2}\sum_{i\in\mathcal{I}^n(I)}\lambda_n\bigg[-\alpha (\prodscal{u^i}{u^{i+1}}+\prodscal{u^{i+1}}{u^{i+2}})+2\prodscal{u^i}{u^{i+2}}\bigg]+\lambda_n\bigg(1+\frac{\alpha^2}{8}\bigg)\#\mathcal{I}^n(I)\notag\\
		& = E_n(u)+\lambda_n\bigg(1+\frac{\alpha^2}{8}\bigg)\#\mathcal{I}^n(I).
	\end{align}
	{Thus we gain a new expression for $\mathcal{E}_n$:}
	\begin{equation}\label{scomposizioneEnergia}
		\mathcal{E}_n= H_n+P_n-\lambda_n\left( 1+ \frac{\alpha^2}{8}\right) \#\mathcal{I}^n(I).
	\end{equation}
	Thanks to this decomposition, we characterize the ground states of $E_n$. 
	\begin{proposition}[{Characterization of the ground states of $\mathcal{E}_n$}]\label{PropMINIMI}
		Let $0 < \alpha \leq 4$. Then, {for $n\in\N$ sufficiently large, it holds
			{\begin{equation*}
					\min_{ u \in  \mathcal{PC}_{\lambda_n}} \mathcal{E}_n(u)= - \lambda_n\#\mathcal{I}^n(I)\bigg[R^2\left( 1+ \frac{ \alpha^2}{8}\right)+(\alpha-1)(1-R^2)\bigg].
			\end{equation*}}
			Furthermore, a minimizer $u_n$ of $\mathcal{E}_n$ over $\mathcal{PC}_{\lambda_n}$  takes values only in one circle $S_\ell$, with $\ell\in\{1,2\}$, and satisfies
			\begin{equation}
				\label{mmm123}
				\prodscal{\pi_{v_\ell^\bot}u_n^i}{\pi_{v_\ell^\bot}u_n^{i+1}}= R^2\frac{\alpha}{4} \quad \text{and}\quad \prodscal{\pi_{v_\ell^\bot}u_n^i}{\pi_{v_\ell^\bot}u_n^{i+2}}= R^2\bigg(\frac{\alpha^2}{8}-1\bigg), \quad \forall i \in \mathcal{I}^n(I).
		\end{equation}}
	\end{proposition}
	\begin{proof}
		{
			Let us postpone the proof of the following equality:
			\begin{equation}\label{Mimim}
				\min_{\substack{u\in\mathcal{PC}_{\lambda_n}\\ u(I)\subset S_1\,\text{or}\,u(I)\subset S_2}}E_n(u)=-\lambda_n\#\mathcal{I}^n(I)\bigg[ R^2\bigg(1+\frac{\alpha^2}{8}\bigg)+(\alpha-1)(1-R^2)\bigg]
			\end{equation}
			after the next claim. We claim that, for $n$ sufficiently large, if $u \in \mathcal{PC}_{\lambda_n}$ is a minimizer of $\mathcal{E}_n$, then $u(I) \subset S_1$ or $u(I) \subset S_2$. We may assume that the open partition associated with $u$ is $\{I_1, I_2\}$, i.e. $M=2$, and $u(I_1) \subset S_1$, $u(I_2) \subset S_2$. The general case $M\in\N$ can be proved similarly.
			We have that
			\begin{equation}\label{14032023formula11}
				\begin{split}
					& \mathcal{E}_n(u)= E_n(u)+ \lambda_n k_nc = \lambda_n\sum_{i\in\mathcal{I}^n(I_1)\cup\mathcal{I}^n(I_2)}\left[ -\alpha\prodscal{u^i}{u^{i+1}}+\prodscal{u^i}{u^{i+2}} \right] \\
					& + \lambda_n\sum_{i\in\mathcal{I}^n(I) \setminus (\mathcal{I}^n(I_1) \cup \mathcal{I}^n(I_2)) ]}\left[ -\alpha\prodscal{u^i}{u^{i+1}}+\prodscal{u^i}{u^{i+2}} \right] + \lambda_n k_n c,
				\end{split}
			\end{equation} 
			where $c:= \vert v_1-v_2 \vert$.
			We observe that $ \# [\mathcal{I}^n(I) \setminus (\mathcal{I}^n(I_1) \cup \mathcal{I}^n(I_2))] \leq 2 $. We define 
			\begin{equation*}
				L:= \min \left\{-\alpha\prodscal{u}{v}+\prodscal{u}{w} -\alpha\prodscal{v}{w}+\prodscal{v}{z} \colon u,v \in S_1,\, w, z \in S_2  \right\} 
			\end{equation*}
			and we observe that 
			\begin{equation*}
				L \leq \sum_{i\in\mathcal{I}^n(I) \setminus (\mathcal{I}^n(I_1) \cup \mathcal{I}^n(I_2))}\left[ -\alpha\prodscal{u^i}{u^{i+1}}+\prodscal{u^i}{u^{i+2}} \right].   
			\end{equation*}
			Therefore by the formula \eqref{14032023formula11} we have that
			\begin{equation}
				\label{eqqn2}
				\min_{\substack{u\in\mathcal{PC}_{\lambda_n}\\ u(I_1)\subset S_1}} E_n(u)+ \min_{\substack{u\in\mathcal{PC}_{\lambda_n}\\ u(I_2)\subset S_2}} E_n(u)+ \lambda_n (L+k_n c) \leq \mathcal{E}_n(u).  
			\end{equation}
			In order to prove the claim we are left to show that, for $n \in \mathbb{N}$ sufficiently large,
			\begin{equation}
				\label{eqqn1}
				\min_{\substack{u\in\mathcal{PC}_{\lambda_n}\\ u(I)\subset S_1\,\text{or}\,u(I)\subset S_2}} E_n(u)< \min_{\substack{u\in\mathcal{PC}_{\lambda_n}\\ u(I_1)\subset S_1}} E_n(u)+ \min_{\substack{u\in\mathcal{PC}_{\lambda_n}\\ u(I_2)\subset S_2}} E_n(u)+ \lambda_n (L+k_n c),
			\end{equation}
			which is equivalent to prove that 
			\begin{multline*}
				-\lambda_n\#\mathcal{I}^n(I)\bigg[ R^2\bigg(1+\frac{\alpha^2}{8}\bigg)+(\alpha-1)(1-R^2)\bigg] \\ < -\lambda_n(\#\mathcal{I}^n(I_1)+\#\mathcal{I}^n(I_2))\bigg[ R^2\bigg(1+\frac{\alpha^2}{8}\bigg)+(\alpha-1)(1-R^2)\bigg]+\lambda_n (L+k_n c),
			\end{multline*}
			where we used formula \eqref{Mimim}.
			Since $ \#\mathcal{I}^n(I)-\# \mathcal{I}^n(I_1)-\# \mathcal{I}^n(I_2) \leq2$, $R\leq 1$ and $ \alpha\leq 4$, we have that, for $n$ sufficiently large,
			\begin{equation*}
				-\lambda_n(\#\mathcal{I}^n(I)-\# \mathcal{I}^n(I_1)-\# \mathcal{I}^n(I_2))\bigg[ R^2\bigg(1+\frac{\alpha^2}{8}\bigg)+(\alpha-1)(1-R^2)\bigg]  \leq 12\lambda_n < \lambda_n(L+k_n c),
			\end{equation*}
			because $k_n \rightarrow + \infty$. We have proved the validity of \eqref{eqqn1}.
			Thus, combining \eqref{eqqn1} and \eqref{eqqn2}, we get
			\begin{equation*}
				\min_{u\in \mathcal{PC}_{\lambda_n}}\mathcal{E}_n(u)=\min_{\substack{u\in\mathcal{PC}_{\lambda_n}\\ u(I)\subset S_1\,\text{or}\,u(I)\subset S_2}}E_n(u).
			\end{equation*}
			We prove that
			\begin{equation*}
				\min_{\substack{u\in\mathcal{PC}_{\lambda_n}\\ u(I)\subset S_1\,\text{or}\,u(I)\subset S_2}}E_n(u)=-\lambda_n\#\mathcal{I}^n(I)\bigg[ R^2\bigg(1+\frac{\alpha^2}{8}\bigg)+(\alpha-1)(1-R^2)\bigg].
			\end{equation*}
			We fix $\ell\in\{1,2\}$ and consider $u\in\mathcal{PC}_{\lambda_n}$ such that $u(I)\subset S_\ell$. By geometric and trigonometric identities we deduce that
			\begin{equation*}
				\prodscal{u^i}{u^{i+1}}=1-R^2+\pi u^i\cdot\pi u^{i+1},
			\end{equation*}
			where $\pi u^i:=\pi_{v_\ell^\bot} u^i$. Of course an analogous statement holds for $\prodscal{u^i}{u^{i+2}}$. Thus
			\begin{align}
				\label{Bb2}
				E_n(u)
				& =\sum_{i\in\mathcal{I}^n(I)}\lambda_n[-\alpha \prodscal{\pi u^i}{\pi u^{i+1}}+\prodscal{\pi u^i}{\pi u^{i+2}}]-(\alpha-1)(1-R^2)\lambda_n\#\mathcal{I}^n(I)\notag\\
				& =\widetilde{E}_n(u)-(\alpha-1)(1-R^2)\lambda_n\#\mathcal{I}^n(I),
			\end{align}
			where we have defined
			\begin{equation*}
				\widetilde{E}_n(u):=\sum_{i\in\mathcal{I}^n(I)}\lambda_n[-\alpha \prodscal{\pi u^i}{\pi u^{i+1}}+\prodscal{\pi u^i}{\pi u^{i+2}}].
			\end{equation*}
			Now we are led to minimize $\widetilde{E}_n$. We find its minimum by following the same argument in \cite{CicaleseSolombrino2015}. With an easy computation similar to the one in \eqref{EspressioneHn}, we remark that
			\begin{align}
				\label{Bb1}
				\widetilde{E}_n(u) 
				& =\frac{1}{2} \lambda_n \sum_{i\in\mathcal{I}^n(I)} \!\!\left \vert \pi u^{i+2}- \frac{\alpha}{2}\pi u^{i+1}+\pi u^{i} \right \vert^{2}\!\!-R^2\bigg(1+\frac{\alpha^2}{8}\bigg)\lambda_n\#\mathcal{I}^n(I)\notag\\
				& = \widetilde{H}_n(u)-R^2\bigg(1+\frac{\alpha^2}{8}\bigg)\lambda_n\#\mathcal{I}^n(I),
			\end{align}
			where 
			\begin{equation*}
				\widetilde{H}_n(u):=\frac{1}{2} \lambda_n \sum_{i\in\mathcal{I}^n(I)} \!\!\left \vert \pi u^{i+2}- \frac{\alpha}{2}\pi u^{i+1}+\pi u^{i} \right \vert^{2}.
			\end{equation*}
			We fix $\phi\in\big[-\frac{\pi}{2},\frac{\pi}{2}\big]$ so that $\cos\phi=\frac{\alpha}{4}$. We may assume for simplicity of notation that $v_\ell=e_n$. Let 
			\begin{equation*}
				u^i:=(R\cos(\phi i),R\sin(\phi i),\sqrt{1-R^2})\in S_\ell,\quad\forall i\in\overline{I}\cap\lambda_n\mathbb Z,
			\end{equation*}
			so that $\pi u^i=(R\cos(\phi i),R\sin(\phi i),0)$. By trigonometric identities, we have that
			\begin{equation*}
				\pi u^i+\pi u^{i+2}=2\cos(\phi)\pi u^{i+1}=\frac{\alpha}{2}\pi u^{i+1},\quad\forall i\in\mathcal{I}^n(I).
			\end{equation*}
			Remarking that $\widetilde{H}_n(u)=0$, we combine the previous identity with \eqref{Bb1} to get that
			\begin{equation*}
				\min_{\substack{u\in\mathcal{PC}_{\lambda_n}\\ u(I)\subset S_1\,\text{or}\,u(I)\subset S_2}}\widetilde{E}_n(u)=-R^2\bigg(1+\frac{\alpha^2}{8}\bigg)\lambda_n\#\mathcal{I}^n(I).
			\end{equation*}
			The computation of the minimum follows from \eqref{Bb2}.\\ 
			\indent Now we consider a minimizer $u\in \mathcal{PC}_{\lambda_n}$ of $\mathcal{E}_n$. For $n$ sufficently large, it must hold that $u(I)\subset S_\ell$, for some $\ell\in\{1,2\}$, and 
			\begin{equation}
				\widetilde{E}_n(u) 
				= -R^2\bigg(1+\frac{\alpha^2}{8}\bigg)\lambda_n\#\mathcal{I}^n(I),
			\end{equation}
			thus implying that $\widetilde{H}_n(u)=0$. It follows that
			\begin{equation*}
				\pi u^{i+1}=\frac{2}{\alpha}\big(\pi u^i+\pi u^{i+2}\big),\quad\forall i\in\mathcal{I}^n(I).
			\end{equation*}
			Squaring the modulus of both sides in the previous equality, we infer
			\begin{equation}
				\prodscal{\pi u^i}{\pi u^{i+2}}=R^2\bigg(\frac{\alpha^2}{8}-1\bigg).
			\end{equation}
			Hence
			\begin{equation*}
				\prodscal{\pi u^i}{\pi u^{i+1}}=\frac{2}{\alpha}\big(\prodscal{\pi u^i}{\pi u^i+\prodscal{\pi u^i}{\pi u^{i+2}}}\big)=\frac{2}{\alpha}\big(R^2+\prodscal{\pi u^i}{\pi u^{i+2}}\big)=R^2\frac{\alpha}{4},
			\end{equation*}
			which concludes the proof.}
	\end{proof}
	{From now on we assume that $n$ is sufficiently large to satisfy the thesis of the above proposition.}
	
	\begin{remark}
		\label{Rem}
		The case $\alpha>4$ is trivial {and} the ground states {of} $\mathcal{E}_n$ are all ferromagnetic, i.e. $u^i = \overline{u}$, for all { $ i\in \overline{I}\cap \lambda_n\mathbb Z$ and for some $\overline{u}\in S_1 \cup S_2$}. {Indeed, denoting by} $ \mathcal{E}_n^{(\alpha=4)}$ the energy of formula \eqref{scomposizioneEnergia} for $\alpha=4$, we have that
		\begin{equation*}
			\mathcal{E}_n(u)= \mathcal{E}_n^{(\alpha=4)}(u)-\lambda_n(\alpha-4) \sum_{i\in\mathcal{I}^n(I)}\prodscal{u^i}{u^{i+1}},
		\end{equation*}
		{for all $ u \in \mathcal{PC}_{\lambda_n}$.}
		By the above proposition, the energy $\mathcal{E}_n^{(\alpha=4)}$  is minimized on ferromagnetic states, which trivially also holds true for the second term in the above sum. The minimal value of ${\mathcal{E}_n}$ is
		\begin{equation*}
			\min_{u \in {\mathcal{PC}_{\lambda_n}}} \mathcal{E}_n(u) = -{\lambda_n}(\alpha-1)\#\mathcal{I}^n(I).
		\end{equation*}
	\end{remark}
	
	\subsection{Zero order $\Gamma$-convergence of $\mathcal{E}_n$}
	\label{GammaConvZero}
	In this subsection we study the $\Gamma$-convergence of $\mathcal{E}_n$  at the zero order. With a slight abuse of notation, we extend the energies $E_n$, $P_n$, $\mathcal{E}_n$ and $H_n$ to the space $\mathrm{L}^\infty(I;\mathrm{co}(S_1)\cup\mathrm{co}(S_2))$, setting their value as $+\infty$ in $\mathrm{L}^\infty(I;\mathrm{co}(S_1)\cup\mathrm{co}(S_2))\setminus\mathcal{PC}_{\lambda_n}$. {With a slight abuse of notation, we extend the projection map $\mathcal{A}$ to the space $\mathrm{L}^\infty(I;\mathrm{co}(S_1)\cup \mathrm{co}(S_2))$ by setting
		\begin{equation}
			\mathcal{A}(u)(t)=
			\begin{cases}
				v_1\quad&\text{if }u(t)\in \mathrm{co}(S_1),\\
				v_2 &\text{if }u(t)\in \mathrm{co}(S_2),
			\end{cases}
		\end{equation}
		for $u\in \mathrm{L}^\infty(I;\mathrm{co}(S_1)\cup \mathrm{co}(S_2))$.
		Furthermore we define}
	\begin{equation}\label{spazfunz1}
		\begin{split}
			\mathfrak{D}
			& :=
			\Big\{u\in\mathrm{L}^\infty(I{;}\,\mathrm{co}(S_1)\cup \mathrm{co}(S_2))\,:\,{\mathcal{A}(u)\in BV(I;\mathrm{co}(S_1)\cup \mathrm{co}(S_2))}\Big\}\\
			& =\mathcal{A}^{-1}(BV(I;\mathrm{co}(S_1)\cup \mathrm{co}(S_2))).
		\end{split}
	\end{equation}
	{It is natural to extend Definition \ref{25052022defC_n} to any spin field $u\in\mathfrak{D}$.} The following {notion of convergence} will be used.
	\begin{definition}\label{Leggediconvergenza}
		Let $ \{u_n \}_{n \in \N} \subset {\mathrm{L}^\infty(I{;}\,\mathrm{co}(S_1)\cup \mathrm{co}(S_2))}$ and $u\in\mathfrak{D}$. We say that $u_n$ ${\mathfrak D}$-converges to $u$ (we write $u_n \overset{{\mathfrak D}}{\rightarrow}u$) if and only if $u_n \weakstar u$ in the weak$^\star$ topology of $\mathrm{L}
		^{\infty}(I;\mathbb{R}^3)$ and $\mathcal{A}(u_n)$ {converges to} $\mathcal{A}(u)$ {weakly$^\star$ in $ BV(I;\{ v_1,v_2 \} )$}.
	\end{definition}
	
	\begin{remark}
		We observe that the {notion of convergence} introduced in the previous definition is induced by the {smallest} topology {on $\mathfrak{D}$} containing the set
		{
			\begin{align*}
				& \big\{A\, \colon A \text{ is an open set of the weak$^\star$ topology of $\mathrm{L}^{\infty}(I;\mathrm{co}(S_1)\cup \mathrm{co}(S_2))$}\\
				& \text{or $A=\mathcal{A}^{-1}(U)$, where $U$ is an open set of the weak$^\star$ topology of $ BV(I;\mathrm{co}(S_1)\cup \mathrm{co}(S_2))$}  \big\}.
			\end{align*}
			For further details about the weak$^\star$ topology of a $BV$ space we address the reader to \cite[Remark 3.12]{AmbFusPal}.
		}
	\end{remark}
	{We prove the following proposition, which relies on the properties contained in Subsection \ref{Some properties} and will be useful in this subsection.}
	\begin{proposition}\label{propcostardueset}
		{Let $\{{u}_n\}_{n \in \mathbb{N}} \subset \mathrm{L}^{\infty}(I;S_1\cup S_2)$ be such that $\mathcal{A}(u_n)\in BV(I;\{v_1,v_2\})$, for any $n\in\N$, and let $\mathcal{C}_n(u_n)=\{ I_j^n\,|\,j\in\{1,\dots,M(u_n)\} \}$ be the open partition associated with $u_n$. We assume} that 
		\begin{equation}
			\label{hphp1}
			\sup_{n \in \mathbb{N}} M({u_n}) <+\infty.
		\end{equation} 
		{Then there exists $u\in\mathfrak{D}$} such that, up to subsequences, ${u}_n {\overset{\mathfrak D}{\rightarrow}} {u}$.
	\end{proposition}
	\begin{proof}
		By Proposition \ref{propLinfliminconv}  {it follows that}, up to a subsequence, $ {u}_n \weakstar {u} \in \mathrm{L}^{\infty}(I; \mathrm{co}({S}_1 \cup {S}_2))$. {Thanks to \eqref{hphp1}, up to the extraction of a subsequence, } {$M=M(u_n)$ is  independent of $n \in \mathbb{N}$. Up to subsequence, $I_j^n\rightarrow I_j$ in the Hausdorff sense, for some intervals $I_j$ and for any $j\in\{1,\dots,M\}$. Note that some $I_j$ could be empty.}
		Let us fix ${j \in \{1,\cdots,M\}}$. For all $\varepsilon>0$ there exists $n_0 \in \mathbb{N}$ such that
		\begin{equation*}
			{(I_j)_\varepsilon=\{t\in I_j\,:\,\mathrm{dist}(t,\dd I_j)>\varepsilon\}  \subset I_{j}^n \quad \forall n\geq n_0. }
		\end{equation*}
		We define the following two sets:
		\begin{align*}
			& A_1= \left\{n \geq n_0 \, : \,{u}_n(t) \in {S}_1 \text{ for a.e. } t \in {(I_j)_\varepsilon}   \right\},\\
			& A_2= \left\{n \geq n_0 \, : \,{u}_n(t) \in {S}_2  \text{ for a.e. } t \in {(I_j)_\varepsilon}    \right\}.
		\end{align*}
		One of the following three {alternatives} may occur:
		\begin{equation*}
			1. \;\; \#A_1= \infty, \; \#A_2< \infty; \quad 2. \; \#A_1< \infty, \;\; \#A_2= \infty; \quad 3. \;\; \#A_1= \infty, \; \#A_2= \infty.  
		\end{equation*}
		In the first case we have that ${u}_n \in \mathrm{L}^{\infty}({(I_j)_\varepsilon}; {S}_1)$ for all $n\geq n_0$, up to {finitely many} indices of the sequence. Thus, by Proposition \ref{propLinfliminconv}, ${u}_n \weakstar {u} \in \mathrm{L}^{\infty}({(I_j)_\varepsilon}; \mathrm{co}({S}_1))$ and hence, by the arbitrariness of $\varepsilon>0$, we obtain that ${u} \in \mathrm{L}^{\infty}({I_j}; \mathrm{co}({S}_1))$. The second case is fully analogous to the first case. If we repeat the above argument for all ${j}\in\{1, {\dots}, M\}$, we {deduce that $u_n\weakstar u$}.\\
		\indent {Finally, we get the thesis by remarking that
			\begin{equation*}
				\lim_{n\rightarrow+\infty}\int_{I}|\mathcal{A}(u_n)-\mathcal{A}(u)|\,dt=\lim_{n\rightarrow+\infty}\sum_{j=1}^M\int_{I_j}|\mathcal{A}(u_n)-\mathcal{A}(u)|\,dt=0.
			\end{equation*}
		}
		\indent {The third alternative leads to a contradiction. Indeed, if it holds true,} we can find two subsequences $ \{n_k^{(1)}\}_{k \in \N}$ and $ \{n_k^{(2)}\}_{k \in \N}$ such that $ {u}_{n_k^{(1)}} \in \mathrm{L}^{\infty}({(I_j)_\varepsilon}; {S}_1)$ and $ {u}_{n_k^{(2)}} \in \mathrm{L}^{\infty}({(I_j)_\varepsilon}; {S}_2)$, for all $k \in \N$. By Proposition \ref{propLinfliminconv}, there exist ${u}_1 \in \mathrm{L}^{\infty}({(I_j)_\varepsilon}; \mathrm{co}({S}_1)) $ and  ${u}_2 \in \mathrm{L}^{\infty}({(I_j)_\varepsilon}; \mathrm{co}({S}_2)) $ such that $ {u}_{n_k^{(1)}} \weakstar {u}_1$ and $ {u}_{n_k^{(2)}} \weakstar {u}_2$. {On} the other hand, {applying again Proposition \ref{propLinfliminconv},} we infer that ${u}_n \weakstar {u} \in \mathrm{L}^{\infty}(I; \mathrm{co}({S}_1 \cup {S}_2))$. Then, by the uniqueness of the limit in the weak${^\star}$ topology, we {infer that $ {u}_{1}(t)={u}_2(t)={u}(t)$} for almost every $t \in {(I_j)_\varepsilon}$, {which is a contradiction since $\mathrm{co}({S}_1) \cap \mathrm{co}({S}_2)= \emptyset$.}\\
	\end{proof}

	Firstly, we study the $\Gamma$-convergence of $E_n$. The following theorem relies on a straightforward application of Theorem \ref{THMaliCicGlor}.\\
	\begin{theorem}\label{TEOasdasd}
		The sequence $E_{n}$ {$\Gamma$-converges} to  the functional
		\begin{equation*}
			\displaystyle
			E(u):=
			\begin{cases}
				\displaystyle \int_{I} f_{hom}(u(t))\,dt & \quad \text{if $u \in \mathrm{L}^{\infty}(I;\mathrm{co}(S_1\cup S_2))$,}\\
				+\infty & \quad \text{otherwise},
			\end{cases}
		\end{equation*}	
		with respect to the {weak$^\star$} topology of $\mathrm{L}^{\infty}(I;\R^3)$, where $f_{hom} \colon \mathrm{co}(S_1\cup S_2) \rightarrow \R$ is defined by
		\begin{equation}\label{fhommme}
			f_{hom}(z)= \lim_{\rho \rightarrow 0} \lim_{k \rightarrow + \infty} \frac{1}{k}  \inf \left\{ \sum_{i=1}^{k-2}\left[ -\alpha\prodscal{u^i}{u^{i+1}}+\prodscal{u^i}{u^{i+2}} \right]\, {\bigg| \,u\text{ s.t. }} (u)_{1,(0,k)}  \in \overline{B(z,\rho)} \right\}.
		\end{equation} 
	\end{theorem}
	\begin{proof}
		The result immediately follows {by} applying Theorem \ref{THMaliCicGlor} to 
		\begin{equation*}
			f^{\xi}(u,v)=
			\begin{cases}
				-\frac{\alpha}{2}\prodscal{u}{v} & \quad \text{if $ \vert \xi \vert=1$},\\
				\frac{1}{2}\prodscal{u}{v} & \quad \text{if $ \vert \xi \vert=2$},
				\\
				0 & \quad \text{otherwise},
			\end{cases}
		\end{equation*}
		where $u,v\in K:=S_1 \cup S_2$, extended to $+\infty$ outside $K$.
	\end{proof}
	\begin{remark}\label{Oss CicSol}
		The function $f_{hom}$ defined in $ \eqref{fhommme} $ does not depend on the parameter $ \lambda_n $. Therefore, in the theorem above the $ \Gamma $-limit does not depend on the choice of $ \lambda_n $.\\
		\indent Furthermore, an analogous statement of Theorem \ref{TEOasdasd} above can be obtained if the functional $E_n$ is defined only in $\mathrm{L}^\infty(I;S_\ell)$ for some $\ell\in\{1,2\}$ (see \cite[Theorem 3.4]{CicaleseSolombrino2015}). {Its} $\Gamma$-limit has the same form and it is finite on $\mathrm{L}^\infty(I;\mathrm{co}(S_\ell))$.
	\end{remark}
	The following theorem is the main result of this subsection.
	\begin{theorem}[{Zero order $\Gamma$-convergence of $\mathcal{E}_n$}]\label{ThmZeroOrder}
		Assume that there exists $\displaystyle\lim_{n \rightarrow + \infty} \lambda_n k_n=:\eta\in (0,+\infty]$. Then the following $\Gamma$-convergence and compactness results hold true.
		\begin{itemize}
			\item[(i)] If $\eta\in(0,+\infty)$, then $\mathcal{E}_n$ {$\Gamma$-converges} to the functional
			\begin{equation*}
				\mathcal{E}(u)=
				\begin{cases}
					\displaystyle \int_{I} f_{hom}(u(t))\,dt +  \eta\vert D \mathcal{A}(u) \vert (I) & \quad \text{if $u \in \mathfrak{D},$}\\
					+\infty & \quad {\text{if $u \in\mathrm{L}^\infty(I;\mathrm{co}(S_1)\cup\mathrm{co}(S_2))\setminus\mathfrak D$}},
				\end{cases}
			\end{equation*}
			with respect to the $\mathfrak{D}$-convergence of Definition \ref{Leggediconvergenza}, where $f_{hom}$ and $\mathfrak{D}$ are defined in \eqref{fhommme} and \eqref{spazfunz1} respectively. Moreover if  $ \{u_n\}_{n \in \N} \subset {\mathrm{L}^\infty(I;\mathrm{co}(S_1)\cup\mathrm{co}(S_2))}$ satisfies
			\begin{equation*}
				\sup_{n  \in \N} \mathcal{E}_{n}(u_n) < +\infty,
			\end{equation*}
			then, up to a subsequence, $u_n \overset{{\mathfrak D}}{\rightarrow} u \in \mathfrak{D}$.
			\item[(ii)] If $\eta=+\infty$, then $\mathcal{E}_n$ {$\Gamma$-converges} to the functional
			\begin{equation*}
				\displaystyle
				\mathcal{E}(u):=
				\begin{cases}
					\displaystyle \int_{I} f_{hom}(u(t))\,dt & \quad \text{if $u \in \mathrm{L}^{\infty}(I;\mathrm{co}(S_1)){\cup} \mathrm{L}^{\infty}(I;\mathrm{co}(S_2)),$ }\\
					+\infty & \quad {\text{if $u\in\mathrm{L}^\infty(I;\mathrm{co}(S_1)\cup\mathrm{co}(S_2))\setminus(\mathrm{L}^{\infty}(I;\mathrm{co}(S_1))\cup \mathrm{L}^{\infty}(I;\mathrm{co}(S_2)))$}},
				\end{cases}
			\end{equation*}
			with respect to the {weak$^\star$} topology of $\mathrm{L}^{\infty}(I;\R^3)$, where $f_{hom}$ is defined in \eqref{fhommme}. Moreover {if $ \{u_n\}_{n \in \N} \subset \mathrm{L}^\infty(I;\mathrm{co}(S_1)\cup\mathrm{co}(S_2))$ satisfies} 
			\begin{equation*}
				\sup_{n  \in \N} \mathcal{E}_{n}(u_n) < +\infty
			\end{equation*}
			then, up to a subsequence, $u_n \weakstar u $ for some $ u \in \mathrm{L}^{\infty}(I;\mathrm{co}(S_1)){\cup} \mathrm{L}^{\infty}(I;\mathrm{co}(S_2))$.
		\end{itemize}

	\end{theorem}
	\begin{proof}
		{We first deal with case (i).} We start {by proving} the compactness result. Let $\{u_n\}_{n \in \N} \subset {\mathrm{L}^\infty(I;\mathrm{co}(S_1)\cup\mathrm{co}(S_2))}$ {be} such that 
		\begin{equation} \label{limitenergiaunif}
			\sup_{n \in \N} \mathcal{E}_n(u_n) <{\overline{C}},
		\end{equation}
		for some ${\overline{C}}>0$. Thus we have that $\{u_n\}_{n\in\N} \subset\mathcal{PC}_{\lambda_n}$. {Let us consider the open partition} $\mathcal{C}_n(u_n)=\{ (I_j)_n\,|\,j\in\{1,\dots,M(u_n)\} \}$ associated with $u_n$, where $M(u_n)-1=\frac{\left\vert D\mathcal{A}(u_n)\right\vert(I)}{\lvert v_1-v_2 \rvert}\in\N$. By formula \eqref{scomposizioneEnergia} and by the definition of $\mathcal{A}$, we compute
		\begin{equation}\label{blablabla}
			\begin{split}
				\mathcal{E}_{n}(u_n)
				& = H_n(u_n)+P_n(u_n)-\lambda_n\left(1+\frac{\alpha^2}{8}\right)\#\mathcal{I}^n(I) \geq P_n(u_n) -\lambda_n\left(1+\frac{\alpha^2}{8}\right)\#\mathcal{I}^n(I) \\
				& =  k_n\lambda_n \left\vert D\mathcal{A}(u_n)\right\vert(I) -\lambda_n\left(1+\frac{\alpha^2}{8}\right)\#\mathcal{I}^n(I)\\
				& = k_n\lambda_n (M(u_n)-1)\vert v_1-v_2 \vert-\lambda_n\left(1+\frac{\alpha^2}{8}\right)\#\mathcal{I}^n(I)\\
				& \geq -C(\alpha)+ k_n\lambda_n (M(u_n)-1) \vert v_1-v_2 \vert,
			\end{split}
		\end{equation}
		for some constant $C=C(\alpha)>0$, where the last inequality is obtained by observing that $\lambda_n\#\mathcal{I}^n(I)= \lambda_n\left \lfloor \frac{1}{\lambda_n}  \right \rfloor-\lambda_n\rightarrow 1$, as $n\rightarrow+\infty$, and thus it is bounded.
		Therefore by {formulae} \eqref{limitenergiaunif} and \eqref{blablabla} we obtain that 
		\begin{equation*}
			\sup_{n \in \N} M(u_n) < C(\eta,{\overline{C}},\alpha,|v_1-v_2|).
		\end{equation*}
		Hence, the sequence $\{u_n\}_{n \in \mathbb{N}} $ satisfies the hypotheses of {Proposition} \ref{propcostardueset} and so we deduce the existence of $u \in \mathfrak{D}$ such that, up to a subsequence, $ u_n \overset{{\mathfrak D}}{\rightarrow} u$. \\
		\indent Now we prove the liminf inequality. Let $\{u_n\}_{n \in \N}\subset{\mathrm{L}^\infty(I;\mathrm{co}(S_1)\cup\mathrm{co}(S_2))}$ {be} such that $ u_n \overset{{\mathfrak D}}{\rightarrow}u \in \mathfrak{D}$. {It is not restrictive to assume that $\{u_n\}_{n \in \mathbb{N}}\subset\mathcal{PC}_{\lambda_n}$}. By the liminf inequality of Theorem \ref{TEOasdasd} we have
		\begin{equation}\label{mumumu1}
			\liminf_{n \rightarrow + \infty} E_n(u_n)\geq \int_{I} f_{hom}(u(t))\,dt.
		\end{equation} 
		On the other hand, by the lower {semicontinuity} of the total variation respect the {weak$^\star$} convergence in $BV(I;\{v_1,\, v_2\})$, we have
		\begin{equation}\label{mumumu2}
			\liminf_{n \rightarrow + \infty} P_{n}(u_n)=\liminf_{n \rightarrow + \infty} k_n \lambda_n \vert D\mathcal{A}(u_n) \vert(I)\geq  \eta \vert D \mathcal{A}(u) \vert(I).
		\end{equation}
		Hence by {formulae} \eqref{mumumu1} and \eqref{mumumu2} we obtain 
		\begin{equation}
			\liminf_{n \rightarrow +\infty} \mathcal{E}_n(u_n) \geq \liminf_{n \rightarrow + \infty} E_n(u_n)+ \liminf_{n \rightarrow + \infty} P_n(u_n) \geq \int_{I}f_{hom}(t)\,dt + \eta \vert D \mathcal{A}(u) \vert.
		\end{equation}
		\indent We finally prove the limsup inequality. Let $u \in \mathrm{L}^{\infty}(I;\mathrm{co}(S_1) \cup \mathrm{co}(S_2))$. We {may} assume that $u\in\mathfrak{D}$. Since $\mathcal{A}(u)\in BV(I;\mathrm{co}(S_1)\cup\mathrm{co}(S_2))$, {it is not restrictive to suppose that the number of jumps of $u$ from one circle to the other is one, i.e. $|D\mathcal{A}(u)|(I)=|v_1-v_2|$}. Furthermore, {by the same density argument exploited in Proposition \ref{propLinfliminconv}} and the locality of the construction, we {may} assume that 
		\begin{equation*}
			u(t)=
			\begin{cases}
				a_1 & \quad \text{if $t \in \big[0,\frac{1}{2}\big]$}, \\
				a_2 & \quad \text{if $ t \in \big(\frac{1}{2},1\big]$},
			\end{cases}
		\end{equation*}
		where $ a_1 \in \mathrm{co}(S_1)$ and $ a_2 \in \mathrm{co}(S_2)$.
		Let $\{ v^j_n\}_{n \in\N}\in\mathrm{L}^\infty(I;S_j)$ be the recovery sequence for the constant function $a_j$ obtained by the $\Gamma$-convergence result in Remark \ref{Oss CicSol} with $2\lambda_n$ as the spacing of the lattice, i.e. {$v^j_n\weakstar a_j$ and}
		\begin{equation}\label{asdasd1234}
			f_{hom}(a_j)=\lim_{n\rightarrow +\infty} E_{n}(v^j_n)=\lim_{n\rightarrow +\infty}2\lambda_n\sum_{i={0}}^{\left \lfloor\frac{1}{2\lambda_n} \right \rfloor-2}\left[ -\alpha\prodscal{(v^j_n)^i}{(v^j_n)^{i+1}}+\prodscal{(v^j_n)^i}{(v^j_n)^{i+2}} \right].
		\end{equation}
		We define
		\begin{equation}
			\label{funzione caso i)}
			u_n(t)=
			\begin{cases}
				v^1_n(2t) & \quad \text{if $t \in \big[0,\frac{1}{2}\big]$}, \\
				v^2_n(2t-1) & \quad \text{if $ t \in \big(\frac{1}{2},1\big]$}.
			\end{cases}
		\end{equation}
		{Remarking that, for all $n\in\N$,
			\begin{equation*}
				\mathcal{A}(u_n)(t)=\mathcal{A}(u)(t)=
				\begin{cases}
					v_1 & \quad \text{if $t \in \big[0,\frac{1}{2}\big]$}, \\
					v_2 & \quad \text{if $ t \in \big(\frac{1}{2},1\big]$},
				\end{cases}
			\end{equation*}
			we deduce that $u_n\overset{{\mathfrak D}}{\rightarrow} u$.} We compute
		\begin{equation}\label{asdasd123}
			\begin{split}
				E_n(u_n)
				&=
				\frac{1}{2}\sum_{i={0}}^{\left \lfloor\frac{1}{2\lambda_n} \right \rfloor-2}2\lambda_n\left[ -\alpha\prodscal{(v^1_n)^i}{(v^1_n)^{i+1}}+\prodscal{(v^1_n)^i}{(v^1_n)^{i+2}} \right]\\
				&+\frac{1}{2}\sum_{i={0}}^{\left \lfloor\frac{1}{2\lambda_n}\right \rfloor-2}2\lambda_n\left[ -\alpha\prodscal{(v^2_n)^i}{(v^2_n)^{i+1}}+\prodscal{(v^2_n)^i}{(v^2_n)^{i+2}} \right]\\
				& + \sum_{i=\left \lfloor\frac{1}{2\lambda_n} \right \rfloor-1}^ {\left \lfloor\frac{1}{2\lambda_n} \right \rfloor}\lambda_n\left[ -\alpha\prodscal{u_n^i}{u_n^{i+1}}+\prodscal{u_n^i}{u_n^{i+2}} \right].
			\end{split}
		\end{equation}
		We observe that 
		\begin{equation}\label{asdasd12}
			\left|\sum_{i=\left \lfloor\frac{1}{2\lambda_n} \right \rfloor-1}^ {\left \lfloor\frac{1}{2\lambda_n} \right \rfloor}\lambda_n\left[ -\alpha\prodscal{u_n^i}{u_n^{i+1}}+\prodscal{u_n^i}{u_n^{i+2}} \right] \right| \leq C(\alpha) \lambda_n \rightarrow 0,
		\end{equation}
		as $n \rightarrow +\infty$.
		By {formulae} \eqref{asdasd1234}, \eqref{asdasd123}, \eqref{asdasd12}, we obtain that 
		\begin{equation}
			\label{convconv1}
			{\lim_{n\rightarrow+\infty}}E_n(u_n){=}\frac{f_{hom}(a_1)+f_{hom}(a_2)}{2}=\int_{I} f_{hom}(u(t))\,dt.\end{equation}
		{Since} $\abs{D\mathcal{A}(u_n)}(I)=\abs{D\mathcal{A}(u)}(I)=\abs{v_1-v_2}$ we get
		\begin{equation}
			\label{convconv4}
			\lim_{n\rightarrow+\infty}P_n(u_n)=\lim_{n\rightarrow+\infty}\lambda_n k_n\abs{v_1-v_2}=\eta\abs{v_1-v_2}.
		\end{equation}
		Combining \eqref{convconv1} and \eqref{convconv4}, we deduce the limsup inequality.\\
		\indent Now we deal with case (ii). {Firstly,} we prove the compactness result. Let $\{u_n\}_{n \in \N} \subset {\mathrm{L}^\infty(I;\mathrm{co}(S_1)\cup\mathrm{co}(S_2))}$ {be} such that 
		\begin{equation*}
			\sup_{n  \in \N} \mathcal{E}_{n}(u_n) <{\overline{C}},
		\end{equation*}
		for some constant ${\overline{C}}>0$. {Thus we have that $\{u_n\}_{n\in\N} \subset\mathcal{PC}_{\lambda_n}$}. With the same compactness argument used in the previous case, we deduce the existence of $ u \in \mathfrak{D}$ such that $u_n \overset{{\mathfrak D}}{\rightarrow} u$. {In particular $u_n\weakstar u$.} {By the lower {semicontinuity} of the total variation respect the {weak$^\star$} convergence in $BV(I;\{v_1,\, v_2\})$, {remarking that $E_n\geq -C(\alpha)$, for some positive constant $C(\alpha)$,} we get
			\begin{equation*}
				\begin{split}
					0=\liminf_{n \rightarrow + \infty} \frac{{\overline{C}}}{\lambda_n k_n}& {\geq}  \liminf_{n \rightarrow + \infty}  \frac{1}{\lambda_n k_n}\left[ E_{n}(u_n)+\lambda_n k_n \vert D \mathcal{A}(u_n) \vert (I) \right]   \\  & \geq \liminf_{n \rightarrow + \infty} \left( {-}\frac{C(\alpha)}{\lambda_n k_n}+ \vert D \mathcal{A}(u_n) \vert (I)\right) \geq \vert D \mathcal{A}(u) \vert (I), 
				\end{split}
			\end{equation*}
		}  
		hence $ u \in \mathrm{L}^{\infty}(I;\mathrm{co}(S_1)) \cup \mathrm{L}^{\infty}(I;\mathrm{co}(S_2)) $. \\
		\indent Let us prove the liminf inequality. {Let $ \{u_n\}_{n \in \N} \subset \mathrm{L}^\infty(I;\mathrm{co}(S_1)\cup\mathrm{co}(S_2))$ be such that $u_n\weakstar u\in\mathrm{L}^\infty(I;\mathrm{co}(S_1)\cup\mathrm{co}(S_2))$} and suppose that 
		\begin{equation*}
			\liminf_{n \rightarrow + \infty} \mathcal{E}_n(u_n) <+\infty.
		\end{equation*} 	
		Up to the extraction of a subsequence, we {may} assume that the previous {lower} limit is actually {a} limit. By compactness, we infer that $ u_n {\weakstar} u\in \mathrm{L}^{\infty}(I; \mathrm{co}(S_1)) \cup \mathrm{L}^{\infty}(I; \mathrm{co} (S_2)) $.
		Hence, by Theorem \ref{TEOasdasd}, we obtain
		\begin{equation*}
			\liminf_{n \rightarrow + \infty}  \mathcal{E}_n(u_n)  \geq \liminf_{n \rightarrow + \infty} E_n(u_n) \geq \int_{I} f_{hom}(u(t))\,dt. 
		\end{equation*}
		\indent We finally prove the limsup inequality. Let $u \in \mathrm{L}^{\infty}(I; \mathrm{co}(S_1)) $, the case $u \in \mathrm{L}^{\infty}(I; \mathrm{co}(S_2)) $ {being} fully analogous. The recovery sequence obtained from Remark \ref{Oss CicSol}, $ \{u_n\}_{n \in \N} \subset \mathrm{L}^{\infty}(I;S_1)$, satisfies the limsup inequality.
	\end{proof}

	\subsection{First order $\Gamma$-convergence of $E_n$}
	\label{Costruzione spin modificati}
	In this subsection and in the following one we study the system when it is close to the helimagnet/ferromagnet transition point as the number of particles diverges. In what follows we {let $\alpha=\alpha_n$ and we assume that $\alpha_n\rightarrow 4^-$, as $n\rightarrow+\infty$, and that $n$ is sufficiently large so that Proposition \ref{PropMINIMI} holds true.}\\
	\indent {Once again, with a slight abuse of notation, we extend the energies $E_n$, $P_n$ and $\mathcal{E}_n$ to the space $\mathrm{L}^\infty(I;\R^3)$, setting their value as $+\infty$ in $\mathrm{L}^\infty(I;\R^3)\setminus\mathcal{PC}_{\lambda_n}$. Similarly, we extend $\mathcal{A}$ from $\mathrm{L}^\infty(I;\mathrm{co}(S_1)\cup\mathrm{co}(S_2))$ to $\mathrm{L}^\infty(I;\R^3)$.}\\
	\indent {The main result of this subsection, Theorem \ref{Teorema Gamma-convergenza ordine 1}, concerns the phenomenon of magnetic anisotropy transitions. Having in mind Proposition \ref{PropMINIMI} and \eqref{Mimim}, we define the functional}
	{\begin{equation*}
			G_n:=E_n- \min_{w \in  \mathcal{PC}_{\lambda_n} }\mathcal{E}_n(w)= E_n- \lambda_n\#\mathcal{I}^n(I)\bigg[R^2\left( 1+ \frac{ \alpha_n^2}{8}\right)+(\alpha_n-1)(1-R^2)\bigg].
	\end{equation*}}
	\indent At this point we need to introduce modified spin {fields} in order to understand better the asymptotic behaviour of the energy {$G_n$}. Let $u \in \mathcal{PC}_{\lambda_n}$ and let $\mathcal{C}_n(u)=\{ {I_j}\,|\,j\in\{1,\dots,M(u)\} \}$ {be} the open partition associated with $u$, with ${I_j=({t_j},t_{j+1})}$, for $j\in\{1,\dots,{M(u)-1}\}$, {and $I_{M(u)}=(t_{M(u)},1)$. We set $t_{M(u)+1}:=\lambda_n\left\lfloor\frac{1}{\lambda_n}\right\rfloor$. Since $u$ is piecewise constant on the edges of the lattice $[0,1]\cap\lambda_n\mathbb Z$, we have that $t_1=0$ and $t_2,\cdots,t_{M(u)+1}$ are multiples of $\lambda_n$, so that $\frac{t_j}{\lambda_n}\in\mathbb N$, for any $j\in\{2,\dots,M(u)+1\}$.}\\
	\indent We define the auxiliary spin $\widetilde{u}_{I_j}\colon \overline{I}_j\rightarrow {S_1\cup S_2}$ {by}
	\begin{equation}\label{funzioneutildaj}
		{\widetilde{u}_{I_j}(t)}=
		\begin{cases}
			{u}(t) \quad & \text{if }t\in {[t_j,t_{j+1})},\\
			w_{j} & \text{if }t={t_{j+1},}
		\end{cases}
	\end{equation}
	{and we set $\widetilde{u}_{I_{M(u)}}(t)=w_{M(u)}$ for $t\in(t_{M(u)+1},1]$}, where $w_j\in {S_1\cup S_2}$ is a vector such that the following boundary condition is satisfied {in $\overline{I}_j$}:
	\begin{equation}
		\label{BDcond}
		\prodscal{u^{ {\frac{t_{j+1}}{\lambda_n}} -1}}{w_j}=\prodscal{u^{{\frac{t_j}{\lambda_n}}}}{u^{ {\frac{t_j}{\lambda_n}} +1}}.
	\end{equation}
	{\indent We prove the following decomposition lemma.
		\begin{lemma}[Decomposition of $G_n$]
			\label{DecompGn}
			Let $u\in\mathcal{PC}_{\lambda_n}$ and let $\mathcal{C}_n(u)=\{I_j\,|\,j\in\{1,\dots,M(u)\} \}$ be the open partition associated with $u$. We have
			\begin{equation}
				\label{DecompositionGn}
				G_n(u)=\sum_{j=1}^{M(u)}MM_n(\widetilde{u}_{I_j})+\sum_{j=1}^{M(u)-1}(R_n)_j(u)+(R_n)_{M(u)}(u)+R_n(u),
			\end{equation}
			where, for all $j\in\{1,\dots,M(u)\}$,
			\begin{align*}
				MM_n(\widetilde{u}_{I_j})
				& :=\lambda_n\sum_{i\in\mathcal{I}^n(I_j)}\left(-\alpha_n \prodscal{\widetilde{u}^{i}_{I_j}}{\widetilde{u}^{i+1}_{I_j}}+\prodscal{\widetilde{u}^{i}_{I_j}}{\widetilde{u}^{i+2}_{I_j}}\right)+\lambda_n{R^2}\left(1+\frac{\alpha_n^2}{8}\right)\#\mathcal{I}^n({I_j})\\
				& +\lambda_n(\alpha_n-1)(1-R^2)\frac{(\#\mathcal{I}^n(I)-M(u)+1)}{M(u)},
			\end{align*}
			and, for all $j\in\{1,\dots,M(u)-1\}$,
			\begin{equation*}
				(R_n)_j(u):=\lambda_n\Big(-\alpha_n \prodscal{u^{\frac{t_{j+1}}{\lambda_n}-1}}{u^{\frac{t_{j+1}}{\lambda_n}}}+\prodscal{u^{\frac{t_{j+1}}{\lambda_n}-1}}{u^{\frac{t_{j+1}}{\lambda_n}+1}}+\prodscal{u^{\frac{t_{j+1}}{\lambda_n}-2}}{u^{\frac{t_{j+1}}{\lambda_n}}}-\prodscal{u^{\frac{t_{j+1}}{\lambda_n}-2}}{w_j}\Big),
			\end{equation*}
			\begin{equation*}
				(R_n)_{M(u)}(u):=\lambda_n\Big(\prodscal{u^{\frac{t_{M(u)+1}}{\lambda_n}-2}}{u^{\frac{t_{M(u)+1}}{\lambda_n}}}-\prodscal{u^{\frac{t_{M(u)+1}}{\lambda_n}-2}}{w_{M(u)}}\Big),
			\end{equation*}
			\begin{equation*}
				R_n(u):=\lambda_n R^2\left( 1+ \frac{ \alpha_n^2}{8}\right)(M(u)-1)+\lambda_n(\alpha_n-1)(1-R^2)(M(u)-1).
			\end{equation*}
		\end{lemma}
		\begin{proof}
			Remarking that
			\begin{equation*}
				\mathcal{I}^n(I_j)=\left\{\frac{t_j}{\lambda_n},\frac{t_j}{\lambda_n}+1,\dots,\frac{t_{j+1}}{\lambda_n}-2\right\},\quad\forall j\in\{1,\dots,M(u)\},
			\end{equation*}
			we may write
			\begin{align*}
				& G_n(u)+\min_{ w \in  \mathcal{PC}_{\lambda_n} }\mathcal{E}_n(w) =\lambda_n \sum_{j=1}^{M(u)-1}\sum_{i\in\mathcal{I}^n(I_j)}\left(-\alpha_n \prodscal{u^i}{u^{i+1}}+\prodscal{u^i}{u^{i+2}}\right)\\
				& +\sum_{j=1}^{M(u)-1}\Big(-\alpha_n\prodscal{u^{\frac{t_{j+1}}{\lambda_n}-1}}{u^{\frac{t_{j+1}}{\lambda_n}}}+\prodscal{u^{\frac{t_{j+1}}{\lambda_n}-1}}{u^{\frac{t_{j+1}}{\lambda_n}+1}}\Big)+\sum_{i\in\mathcal{I}^n(I_{M(u)})}\left(-\alpha_n \prodscal{u^i}{u^{i+1}}+\prodscal{u^i}{u^{i+2}}\right).\\
			\end{align*}
			After adding and subtracting the terms $\prodscal{u^{\frac{t_{j+1}}{\lambda_n}-2}}{w_j}$, for any $j\in\{1,\dots,M(u)\}$, we interchange $\prodscal{u^{\frac{t_{j+1}}{\lambda_n}-2}}{w_j}$ and $\prodscal{u^{\frac{t_{j+1}}{\lambda_n}-2}}{u^{\frac{t_{j+1}}{\lambda_n}}}$ in the first and the third sums, for any $j\in\{1,\dots,M(u)\}$, obtaining
			\begin{align*}
				& G_n(u)+\min_{w\in  \mathcal{PC}_{\lambda_n}}\mathcal{E}_n(w)\\
				& = \lambda_n\Bigg[\sum_{j=1}^{M(u)}\sum_{i\in\mathcal{I}^n(I_j)}\left(-\alpha_n \prodscal{\widetilde{u}_{I_j}^i}{\widetilde{u}_{I_j}^{i+1}}+\prodscal{\widetilde{u}_{I_j}^i}{\widetilde{u}_{I_j}^{i+2}}\right)\\
				& +\sum_{j=1}^{M(u)-1}\Big(-\alpha_n \prodscal{u^{\frac{t_{j+1}}{\lambda_n}-1}}{u^{\frac{t_{j+1}}{\lambda_n}}}+\prodscal{u^{\frac{t_{j+1}}{\lambda_n}-1}}{u^{\frac{t_{j+1}}{\lambda_n}+1}}+\prodscal{u^{\frac{t_{j+1}}{\lambda_n}-2}}{u^{\frac{t_{j+1}}{\lambda_n}}}-\prodscal{u^{\frac{t_{j+1}}{\lambda_n}-2}}{w_j}\Big)\\
				& + \Big(\prodscal{u^{\frac{t_{M(u)+1}}{\lambda_n}-2}}{u^{\frac{t_{M(u)+1}}{\lambda_n}}}-\prodscal{u^{\frac{t_{M(u)+1}}{\lambda_n}-2}}{w_{M(u)}}\Big)\Bigg]\\
				& = \sum_{j=1}^{M(u)}MM_n(\widetilde{u}_{I_j})+\sum_{j=1}^{M(u)-1}(R_n)_j(u)+(R_n)_{M(u)}(u)-\lambda_n R^2\bigg(1+\frac{\alpha_n^2}{8}\bigg)\sum_{j=1}^{M(u)}\#\mathcal{I}^n(I_j)\\
				& -\lambda_n(\alpha_n-1)(1-R^2)(\#\mathcal{I}^n(I)-M(u)+1).
			\end{align*}
			We conclude the proof by computing
			\begin{align*}
				& -\min_{ w \in  \mathcal{PC}_{\lambda_n}}\mathcal{E}_n(w)-\lambda_nR^2\bigg(1+\frac{\alpha_n^2}{8}\bigg)\sum_{j=1}^{M(u)}\#\mathcal{I}^n(I_j)\\
				& -\lambda_n(\alpha_n-1)(1-R^2)(\#\mathcal{I}^n(I)-M(u)+1)\\
				& = \lambda_n R^2\left( 1+ \frac{ \alpha_n^2}{8}\right)\bigg[\#\mathcal{I}^n(I)-\sum_{j=1}^{M(u)}\#\mathcal{I}^n(I_j)\bigg]+\lambda_n(\alpha_n-1)(1-R^2)(M(u)-1)\\
				& =\lambda_n R^2\left( 1+ \frac{ \alpha_n^2}{8}\right)(M(u)-1)+\lambda_n(\alpha_n-1)(1-R^2)(M(u)-1)=R_n(u),
			\end{align*}
			where we used
			\begin{equation}\label{ContoIndici}
				\sum_{j=1}^M\#\mathcal{I}^n(I_j)=\#\mathcal{I}^n(I)-M(u){+}1.
			\end{equation}
		\end{proof}
	}
	{\begin{remark}
			In the decomposition \eqref{DecompositionGn} of $G_n(u)$ the functional $MM_n(\widetilde{u}_{I_j})$ represents the energy of the $j$-th modified spin field $\widetilde{u}_{I_j}$,  which is localized in one circle. The remainders for such modifications, $(R_n)_j(u)$ and $(R_n)_{M(u)}(u)$, consist of the interactions between spins with values in two neighboring intervals, $I_j$ and $I_{j+1}$. Furthermore, they contain an additional term linked to the boundary condition \eqref{BDcond}. The term $R_n(u)$ contains a corrective addend.
	\end{remark}}
	\begin{remark}\label{MM_nèpositivo}
		{Following the same computations done in \eqref{EspressioneHn}, we infer that $MM_n(\widetilde{u}_{I_j}) \geq 0$, for all $j\in\{1, \dots,\, M(u)\}$ and $u \in \mathcal{PC}_{\lambda_n}$.}\\
	\end{remark}
	{The next theorem shows that the correct scaling of the energy to capture spin fields' magnetic anisotropy transitions is $\lambda_n$. To this end, for $M\in\N$, we set
		\begin{align*}
			\mathfrak R_M:=\inf 
			& \bigg\{ \liminf_{n \rightarrow + \infty} \frac{1}{\lambda_n}\Bigg[\sum_{j=1}^{M-1}(R_n)_j(u_n)+(R_n)_M(u_n)+R_n(u_n)\Bigg]\bigg|\{u_n\}_{n\in\N}\subset\mathcal{PC}_{\lambda_n}\text{ such that }\\
			& \mathcal{A}(u_n) \weakstarBV v\in BV(I;\{v_1,v_2\}),\text{ with }M=\frac{|D v|(I)}{|v_1-v_2|}+1\in\N\bigg\}.
		\end{align*}
	}
	\begin{theorem}[{First order $\Gamma$-convergence of $E_n$}]
		\label{Teorema Gamma-convergenza ordine 1}
		Assume that there exists $\displaystyle\lim_{n \rightarrow + \infty} \lambda_n k_n=:\eta \in (0,+\infty)$. Then the following compactness and $\Gamma$-convergence results hold true:
		\begin{itemize}
			\item[(i)](Compactness) If for $\{u_n\}_{n \in \N} \subset \mathrm{L}^{\infty}(I;\R^3)$ there exists a {constant} $C>0$ independent of $n$ such that
			\begin{equation}\label{compEnglimitataordi1}
				{\sup_{n\in\N}G_n(u_n)\leq \lambda_n C\quad \text{and}\quad \sup_{n\in\N}P_n(u_n) \leq C,}
			\end{equation}
			then, up to {subsequences}, $\mathcal{A}(u_n) {\weakstarBV} v \in  BV(I;\{v_1, v_2\})$.
			\item[(ii)](liminf inequality) For all $ v\in BV(I;\{ v_1, v_2\})$ and $ \{ u_n\}_{n \in \N}\subset \mathcal{PC}_{\lambda_n}$ such that $\mathcal{A}(u_n){\weakstarBV} v$ and {\eqref{compEnglimitataordi1} holds} for some {constant} $C >0$, then
			\begin{equation*}
				\liminf_{n \rightarrow + \infty} \frac{{G_n}(u_n)}{\lambda_n} \geq {\mathfrak R_M},
			\end{equation*}
			{where $M=\frac{|D v|(I)}{|v_1-v_2|}+1\in\N$.}
			\item[(iii)](limsup inequality) For all $ v \in BV(I; \{ v_1, v_2\})$ there exists $\{u_n\}_{n \in N}\subset \mathcal{PC}_{\lambda_n}$ such that $\mathcal{A}(u_n){\weakstarBV} v$, {\eqref{compEnglimitataordi1} holds} for some {constant} $C >0$ {and}
			\begin{equation*}
				\lim_{n \rightarrow + \infty} \frac{{G_n}(u_n)}{\lambda_n} = {\mathfrak R_M}, 
			\end{equation*}
			{where $M=\frac{|D v|(I)}{|v_1-v_2|}+1\in\N$.}
		\end{itemize}
	\end{theorem}
	\begin{proof}	
		We start {by} proving (i). {Let $\{u_n\}_{n\in\N}\subset\mathrm{L}^\infty(I;\R^3)$ be such that \eqref{compEnglimitataordi1} holds true. It follows that $\{u_n\}_{n\in\N}\subset\mathcal{PC}_{\lambda_n}$.} Since $\eta\in(0,+\infty)$, by the second inequality of formula \eqref{compEnglimitataordi1}, we deduce that the sequence $\{|D\mathcal{A}(u_n)|(I)\}_{n\in\N}$ is bounded and so the sequence $\{\mathcal{A}(u_n)\}_{n \in \N}$ is bounded in the space $BV(I;\{ v_1, v_2\})$. Thus, up to {a} subsequence, $\{\mathcal{A}(u_n)\}_{n \in \N}$ converges to a function $v \in BV(I; \{ v_1, v_2\})$ {weakly$^\star$ in} $BV(I; \{v_1, v_2\})$. \\
		\indent We prove (ii). Let {$ v\in BV(I;\{ v_1, v_2\})$ and $ \{ u_n\}_{n \in \N}\subset \mathcal{PC}_{\lambda_n}$ {be} such that $\mathcal{A}(u_n)\weakstarBV v$ and \eqref{compEnglimitataordi1} holds}. By assumption, $\{|D\mathcal{A}(u_n)|(I)\}_{n\in\N}$ is bounded. Let $\mathcal{C}_n(u_n)=\{ (I_{j})_n\,|\,j\in\{1,\dots,M(u_n)\} \}$ {be} the open partition associated with $u_n$. {Up to subsequences, we may assume that $M=M(u_n)$ is independent of $n$.}
		By Lemma \ref{DecompGn}, Remark \ref{MM_nèpositivo} and the definition of ${\mathfrak R_M}$ we have
		\begin{align*} 
			& \liminf_{n \rightarrow + \infty} \frac{{G_n}(u_n)}{\lambda_n}\\
			& \geq \liminf_{n \rightarrow + \infty} \sum_{j=1}^{M} \frac{ MM_n({\widetilde{u}_n}{_{(I_j)_n})}}{\lambda_n} + \liminf_{n \rightarrow + \infty}{\Bigg[\sum_{j=1}^{M-1}\frac{(R_n)_j(u_n)}{\lambda_n}+\frac{(R_n)_{M}(u_n)+R_n(u_n)}{\lambda_n}\Bigg]} \\
			& \geq \liminf_{n \rightarrow + \infty} {\Bigg[\sum_{j=1}^{M-1}\frac{(R_n)_j(u_n)}{\lambda_n}+\frac{(R_n)_{M}(u_n)+R_n(u_n)}{\lambda_n}\Bigg]\geq\mathfrak R_M.}
		\end{align*}
		\indent We finally prove (iii). {Let $ v \in BV(I; \{ v_1, v_2\})$.} {It is not restrictive to assume that $v=v_1 \chi_{\left[0,\frac{1}{2}\right]} +  v_2 \chi_{\left(\frac{1}{2},1\right]}$ and thus we can choose $u\in\mathcal{PC}_{\lambda_n}$ such that $\mathcal{A}(u)=v$}. By the definition of ${\mathfrak R_M}$ and by \cite[Theorem 4.2]{CicaleseSolombrino2015}, we gain the existence of $\{u_n\}_{n\in\N}$ such that $\mathcal{A}(u_n) {\weakstarBV}\mathcal{A}(u)$, $u_n \chi_{\left[0,\frac{1}{2}\right]} \in S_1, \; \;  u_n \chi_{\left(\frac{1}{2},1\right]} \in S_2$ and the following formulae are satisfied:
		\begin{equation*}
			{\lim_{n\rightarrow+\infty}\frac{1}{\lambda_n}\Bigg[\sum_{j=1}^{M-1}(R_n)_j(u_n)+(R_n)_M(u_n)+R_n(u_n)\Bigg]=\mathfrak R_M},
		\end{equation*}
		\begin{equation*}          {\bigg(1-\frac{\alpha_n}{4}\bigg)^{-\frac{3}{2}}}\frac{MM_n\big(u_{n}\chi_{\left[0,\frac{1}{2}\right]}\big)}{\lambda_n }{<C},\quad {\bigg(1-\frac{\alpha_n}{4}\bigg)^{-\frac{3}{2}}}\frac{MM_n\big(u_{n}\chi_{\left(\frac{1}{2},1\right]}\big) }{\lambda_n } <C.   
		\end{equation*}
		Therefore 
		\begin{equation*}
			\lim_{n \rightarrow + \infty} \frac{{G_n}(u_n)}{\lambda_n}={\mathfrak R_M}. 
		\end{equation*}
	\end{proof}
	\subsection{Second order $\Gamma$-convergence of $E_n$}
	$\\ $
	{We {let} $ \alpha=\alpha_n:= 4(1-\delta_n)$, where $ \{\delta_n\}_{n\in\N}$ is a positive vanishing sequence.}\\
	\indent At the second order we split the global functional on the {2}-dimensional sphere {into finitely many} functionals localized in {circles}, where we repeat the analysis lead in \cite{CicaleseSolombrino2015}. For each circle $S_\ell$ we define a convenient order parameter.\\  
	\indent Let $u \in \mathcal{PC}_{\lambda_n}$. {According to the notation introduced in Subsection \ref{Costruzione spin modificati}, for $j\in\{1,\dots,M(u)\}$ and $i\in\left\{\frac{t_j}{\lambda_n},\frac{t_j}{\lambda_n}+1,\dots,\frac{t_{j+1}}{\lambda_n}-1\right\}$, we consider the pair $\big(\widetilde{u}^i_{I_j}, \widetilde{u}^{i+1}_{I_j}\big)$ of vectors that take values in $S_\ell$, for some $\ell=\ell_j\in\{1,2\}$. We associate each pair} with the corresponding oriented angle ${\theta_{I_j}^i}\in[-\pi,\pi)$ with vertex {in} the center of the {circle $S_\ell$} given by
	\begin{equation*}
		{\theta_{I_j}^i:=\chi\left[\pi_{v_\ell^\bot}(\widetilde{u}^i_{I_j}),\pi_{v_\ell^\bot}(\widetilde{u}^{i+1}_{I_j})\right]\arccos\Big({\prodscal{\pi_{v_\ell^\bot}(\widetilde{u}^i_{I_j})}{\pi_{v_\ell^\bot}(\widetilde{u}^{i+1}_{I_j})}}\Big).}
	\end{equation*}
	We set
	\begin{equation}
		\label{w}
		w_{I_j}^{i}:={\sqrt{{\frac{8}{4-\alpha_n}}}\sin{\frac{\theta_{I_j}^i}{2}}}=\sqrt{\frac{2}{\delta_n}}\sin{\frac{\theta_{I_j}^i}{2}}
	\end{equation}
	and
	\begin{equation*}
		w(t)=w_{I_j}^{i}\quad\text{{for} }t\in\lambda_n\{i+[0,1)\},\,i\in\left\{\frac{t_j}{\lambda_n},\dots,\frac{t_{j+1}}{\lambda_n}-1\right\},\,j\in\{1,\dots,{M(u)}\}.
	\end{equation*}
	{We extend $w(t)=w^{\frac{t_{j+1}}{\lambda_n}-1}_{I_{M(u)}}$, for $t\in[t_{M(u)+1},1]$, so that $w$ is well-defined in the whole interval {$\overline{I}$.}} Note that we can define a map {$T_n$ by setting}
	\begin{equation*}
		{T_n(u):=(w, \mathcal{A}(u))},\quad\forall u\in\mathcal{PC}_{\lambda_n},
	\end{equation*}
	and we denote $\widetilde{\mathcal{PC}}_{\lambda_n}:=T_n(\mathcal{PC}_{\lambda_n})$. We observe that if $ h= T_n(u)=T_n(v)$ then $u(t)$ and $v(t)$ {belong to the same circle, for any $t\in \overline{I}$,} and $u,v$ differ by a constant rotation. Furthermore, $ {G_n}(u)={G_n}(v)$ and $P_n(u)=P_n(v)$. The same identity holds for the functionals defined in Lemma \ref{DecompGn}.
	Therefore, with a slight abuse of notation, we now {set}
	\begin{equation}\label{NEWH_n^hf}
		{G_n}(h):=
		\begin{cases*}
			{G_n}(u) & \text{if $h \in \widetilde{\mathcal{PC}}_{\lambda_n}$},\\
			+\infty & \text{otherwise,}
		\end{cases*}
	\end{equation}
	\begin{equation}
		\label{NEWP_n}
		P_{n}(h):=
		\begin{cases*}
			P_n(u) & \text{if $h \in \widetilde{\mathcal{PC}}_{\lambda_n}$},\\
			+\infty & \text{otherwise},
		\end{cases*}
	\end{equation}
	\begin{equation}
		\label{NEWMM_n}
		MM_{n}(h_{|I_j}):=
		\begin{cases*}
			MM_n(\widetilde{u}_{I_j}) & \text{if $h \in \widetilde{\mathcal{PC}}_{\lambda_n}$},\\
			+\infty & \text{otherwise},
		\end{cases*}
	\end{equation}
	\begin{equation}\label{NEWR_nj}
		(R_n)_j(h):=
		\begin{cases*}
			{(R_n)_j}(u) & \text{if $h \in \widetilde{\mathcal{PC}}_{\lambda_n}$},\\
			+\infty & \text{otherwise,}
		\end{cases*}
	\end{equation}
	{
		\begin{equation}\label{NEWR_nMu}
			(R_n)_{M(u)}(h):=
			\begin{cases*}
				(R_n)_{M(u)}(u) & \text{if $h \in \widetilde{\mathcal{PC}}_{\lambda_n}$},\\
				+\infty & \text{otherwise,}
			\end{cases*}
		\end{equation}
		\begin{equation}\label{NEWR_n}
			R_n(h):=
			\begin{cases*}
				R_n(u) & \text{if $h \in \widetilde{\mathcal{PC}}_{\lambda_n}$},\\
				+\infty & \text{otherwise,}
			\end{cases*}
		\end{equation}
	}
	for $j\in\{ 1,\dots,M(h)\}$, where ${h \in \mathrm{L}^1(I;\R\times \{v_1,v_2\})}$, $h=T_n(u)$ and $M(h):=M(u)$.
	
	We want to study the convergence of the functional {
		\begin{align*}
			\label{FunzFinale}
			\mathcal{G}_n(h)
			& =
			\begin{cases}   
				\displaystyle G_n(h)-\sum_{j=1}^{M(h)-1}(R_n)_j(h)+(R_n)_{M(h)}(h)+R_n(h)\quad&\text{if }h\in\widetilde{\mathcal{PC}}_{\lambda_n},\\
				+\infty &\text{otherwise},            \end{cases}\\
			& =\sum_{j=1}^{M(h)}MM_n(\widetilde{h}_{|I_j})
		\end{align*}
	}
	for $h\in\mathrm{L}^1(I;\R\times\{v_1,v_2\})$. In order to establish the related result, we need a notion of convergence.
	
	\begin{definition}
		\label{ConvergenzaLtheta}
		Let $ \{h_n\}_{n\in\N} \subset \widetilde{\mathcal{PC}}_{\lambda_n}$ and $h\in\mathrm{L}^{1}(I;\R\times \{v_1, v_2\})$. We say that $h_n$ ${\theta}$-converges to $h$ (we write $h_n \Ltheta h $) if and only if the following conditions are satisfied:
		\begin{itemize}
			\item there exist $\{u_n\}_{n\in\N}\subset\mathcal{PC}_{\lambda_n}$ and a positive constant $C$ such that if $\mathcal{C}_n(u_n)=\{ (I_{j})_n\,|\,j\in\{1,\dots,M(u_n)\} \}$ is the open partition associated with $u_n$, {then}
			\begin{itemize}
				\item $h_n=T_n(u_n)$ and $P_n(h_n)<C$,
				\item $M(u_n)\rightarrow M\in\N$ as $n\rightarrow+\infty$,
				\item $(I_j)_n\rightarrow I_{j}$ in the Hausdorff sense, as $n\rightarrow +\infty$, for any $j\in\{1,\dots,M\}$.
			\end{itemize}
			
			\item ${h_n}\chi_{(I_j)_n}\rightarrow h\chi_{I_j}$ in $\mathrm{L}^1(I;\R\times\{v_1{,v_2}\})$, for all $j\in\{1,\dots,M\}$.
		\end{itemize}
	\end{definition}
	{We point out that the intervals $I_j$ of the previous definition may be also empty.\\
		\indent The next theorem shows that the correct scaling of the energy to capture spin fields' chirality transitions is $\sqrt{2}\lambda_n\delta_n^{\frac{3}{2}}$.}
	\begin{theorem}[{Second order $\Gamma$-convergence of $E_n$}]
		\label{Teoremafinale}
		Assume that there exist $\displaystyle\lim_{n \rightarrow + \infty} \lambda_n k_n=:\eta \in (0,+\infty)$ and $\displaystyle l:=\lim_{n \rightarrow + \infty}\frac{\lambda_n}{(2\delta_n)^{\frac{1}{2}}}\in[0,+\infty]$. 
		Then the following statements are true:
		
		\begin{itemize}
			\item[(i)](Compactness) If for $\{h_n\}_{n \in \N}\subset \mathrm{L}^{1}(I;\R \times \{ v_1,v_2\})$ there exists a constant $C>0$ such that
			\begin{equation}\label{compEnglimitataordi12}
				{\sup_{n\in\N}\mathcal{G}}_n(h_n)\leq \sqrt{2}\lambda_n\delta_n^{\frac{3}{2}} C \quad\text{and}\quad {\sup_{n\in\N}}P_n(h_n)\leq C,
			\end{equation}
			then, up to a subsequence, 
			$h_n\Ltheta h$, where
			\begin{itemize}
				\item if $l=0$, $h
				\in  BV(I;	\{-1,1\} \times \{v_1, v_2\})$;
				\item if $l \in(0,+\infty)$, $ h_{| I_j} \in \mathrm{H}_{| per| }^1(I_j;\R\times \{ v_1,v_2 \})$ for all $j\in\{1,\dots,M(h)\}$;
				\item if $l = + \infty $, $h$ is piecewise constant with values in $\R \times\{v_1,v_2\}$.
			\end{itemize}
			The space $ \mathrm{H}_{|per| }^{1}((a,b);\R \times \{v_1,  v_2\})$ is equal to  
			\begin{equation*}
				\left\{ h \in H^1((a,b);\R  \times \{v_1, v_2\}): \, \vert w(a) \vert= \vert w(b) \vert \text{ where $h=(w,\mathcal{A}(u))$} \right\}.
			\end{equation*}
			\item[(ii)](liminf inequality)
			\begin{itemize}
				\item If $l=0$, for all $h=(w,\mathcal{A}(u))
				\in  BV(I;\{-1,1\} \times \{v_1,v_2\})$ and for all $ \{ h_n\}_{n \in \N}\subset \widetilde{PC}_{\lambda_n}$ such that $ h_n \Ltheta h$
				and {\eqref{compEnglimitataordi12} holds true
					for some constant $C >0$}, then
				\begin{equation*}
					\liminf_{n \rightarrow + \infty} \frac{{\mathcal{G}_n}(h_n)}{\sqrt{2}\lambda_n\delta_n^{\frac{3}{2}}} \geq \frac{{4}}{3} {R^2}\sum_{j=1}^{M(h)}\left| Dw \right|(I_j).
				\end{equation*}
				\item If $l \in(0,+\infty)$, for all $h=(w,\mathcal{A}(u))\in \mathrm{L}^1(I;\R \times  \{v_1,v_2\})$ such that $h_{| I_j} \in \mathrm{H}_{| per| }^1(I_j;\R \times \{v_1, v_2\})$, for {every} $j\in\{1,\dots,M(h)\}$, and for all $ \{ h_n\}_{n \in \N}\subset \widetilde{PC}_{\lambda_n}$ such that $h_n \Ltheta h$ 
				and {\eqref{compEnglimitataordi12} holds true for some constant $C>0$,}	then
				\begin{equation*}
					\liminf_{n \rightarrow + \infty} \frac{{\mathcal{G}_n}(h_n)}{\sqrt{2}\lambda_n\delta_n^{\frac{3}{2}}} \geq  \sum_{j=1}^{M(h)}{R^2} \left[\frac{1}{l} \int_{I_j}(w^2(x)-1)^2\,dx+l \int_{I_j} (w'(x))^2\,dx \right].
				\end{equation*}
				
				\item If $l=+\infty$, for all {piecewise constant functions $h\colon I\rightarrow \R\times \{v_1, v_2\}$}  and for all $ \{ h_n\}_{n \in \N}\subset \widetilde{PC}_{\lambda_n}$ such that 
				$ h_n \Ltheta h,$
				and {\eqref{compEnglimitataordi12} holds true
					for some constant $C >0$},
				then
				\begin{equation*}
					\liminf_{n \rightarrow + \infty} \frac{{\mathcal{G}_n}(h_n)}{\sqrt{2}\lambda_n\delta_n^{\frac{3}{2}}} \geq 0.
				\end{equation*}
			\end{itemize}
			
			\item[(iii)](limsup inequality)
			\begin{itemize}
				\item If $l=0$, for all $ h=(w,\mathcal{A}(u)) \in BV(I; 	\{-1, 1\} \times \{v_1,v_2\})$ there exists $\{h_n\}_{n \in N}\subset \widetilde{\mathcal{PC}}_{\lambda_n}$ such that $ h_n \Ltheta h$, {\eqref{compEnglimitataordi12} holds true
					for some constant $C >0$} and
				\begin{equation*}
					\lim_{n \rightarrow + \infty}\frac{{\mathcal{G}_n}(h_n)}{\sqrt{2}\lambda_n\delta_n^{\frac{3}{2}}} = \frac{{4}}{3} {R^2}\sum_{j=1}^{M}\left| D w \right|( I_j).  
				\end{equation*}
				
				\item If $l\in(0,+\infty)$, for all $h=(w,\mathcal{A}(u))\in \mathrm{L}^1(I;\R \times  \{v_1,  v_2\})$ such that $h_{| I_j} \in \mathrm{H}_{| per| }^1(I_j; \R \times \{v_1, v_2\})$ for all $j\in\{1,\dots,M(h)\}$, there exists $\{h_n\}_{n \in N}\subset \widetilde{\mathcal{PC}}_{\lambda_n}$ such that $ h_n \Ltheta h,$ {\eqref{compEnglimitataordi12} holds true
					for some constant $C >0$} and
				\begin{equation*}
					\lim_{n \rightarrow + \infty}\frac{{\mathcal{G}_n}(h_n)}{\sqrt{2}\lambda_n\delta_n^{\frac{3}{2}}} = \sum_{j=1}^{M} {R^2}\left[\frac{1}{l} \int_{I_j}(w^2(x)-1)^2\,dx+l \int_{I_j} (w'(x))^2\,dx \right].    
				\end{equation*}
				
				\item If $l=+\infty$, for all {piecewise constant functions $h\colon I\rightarrow \R\times \{v_1, v_2\}$} there exists $ \{ h_n\}_{n \in \N}\subset \widetilde{\mathcal{PC}}_{\lambda_n}$ such that $ h_n \Ltheta h,$
				{\eqref{compEnglimitataordi12} holds true
					for some constant $C >0$} and
				\begin{equation*}
					{\lim_{n \rightarrow + \infty}} \frac{{\mathcal{G}_n}(h_n)}{\sqrt{2}\lambda_n\delta_n^{\frac{3}{2}}} = 0.
				\end{equation*}
			\end{itemize}
		\end{itemize}
	\end{theorem}
	\begin{proof}
		We prove the statement only in the case $ l=0 $, the other cases {being} fully analogous.
		We start {by} proving (i). {Let $\{h_n\}_{n \in \N}\subset \mathrm{L}^{1}(I;\R \times \{ v_1,v_2\})$ be such that \eqref{compEnglimitataordi12} holds true for some constant $C >0$}. By {formula and Remark \ref{MM_nèpositivo},} we infer that
		\begin{equation}\label{Mainthmformula11}
			MM_{n}(h_{n| I_j^n}) \leq \lambda_n \delta_n^{\frac{3}{2}} C, \quad \text{for all $j\in \{ 1, \dots, M(h_n)\}$ and $n\in \N$.}
		\end{equation}
		\indent It is easy to see that, up to subsequences, $M=M(h_n)$ is independent of $n \in \N$ and the interval $(I_j)_n \rightarrow I_j=(t_{j-1},\, t_j)$, in the Hausdorff sense, for every $j\in\{1,\dots,M(h_n)$\} (it may happen that $I_j=\emptyset$, for some $j$). In the following computations we drop for simplicity the dependence on $n$ writing $I_j$ in place of $(I_j)_n$.\\
		\indent {Reasoning as in Proposition \ref{PropMINIMI}, thanks to \eqref{ContoIndici}, we compute
			\begin{align*}
				& \mathcal{G}_n(h_n)
				=\sum_{j=1}^M\lambda_n\sum_{i\in\mathcal{I}^n(I_j)}\left(-\alpha_n \prodscal{\pi\widetilde{u}^{i}_{nI_j}}{\pi\widetilde{u}^{i+1}_{nI_j}}+\prodscal{\pi\widetilde{u}^{i}_{nI_j}}{\pi\widetilde{u}^{i+2}_{nI_j}}\right)+\lambda_n{R^2}\left(1+\frac{\alpha_n^2}{8}\right)\sum_{j=1}^M\#\mathcal{I}^n({I_j})\\
				& +\lambda_n(\alpha_n-1)(1-R^2)(\#\mathcal{I}^n(I)-M+1)-\lambda_n(\alpha_n-1)(1-R^2)\sum_{j=1}^M\#\mathcal{I}^n(I_j)\\
				& =\sum_{j=1}^M\lambda_n\sum_{i\in\mathcal{I}^n(I_j)}\left(-\alpha_n \prodscal{\pi\widetilde{u}^{i}_{nI_j}}{\pi\widetilde{u}^{i+1}_{nI_j}}+\prodscal{\pi\widetilde{u}^{i}_{nI_j}}{\pi\widetilde{u}^{i+2}_{nI_j}}\right)+\lambda_n{R^2}\left(1+\frac{\alpha_n^2}{8}\right)(\#\mathcal{I}^n(I)-M+1),
			\end{align*}
			where we set $\pi\widetilde{u}_{nI_j}^i:=\pi_{v_\ell^\bot}\widetilde{u}_{nI_j}^i$, with $\ell=\ell_j\in\{1,2\}$ such that $\widetilde{u}_{nI_j}^i\in S_\ell.$}
		
		By the definition of $\widetilde{u}^{i}_{nI_j}$ {and geometric and trigonometric identities,} we {observe} that
		\begin{equation*}
			{R^2-\prodscal{\pi\widetilde{u}^i_{nI_j}}{\pi\widetilde{u}^{i+1}_{nI_j}}}=2R^2\sin^2\bigg( \frac{\theta^i_{I_j}}{2} \bigg),
		\end{equation*}
		\begin{equation*}
			{R^2-\prodscal{\pi\widetilde{u}^i_{nI_j}}{\pi\widetilde{u}^{i+2}_{nI_j}}}=R^2[1-\cos(\theta^i_{I_j}+\theta^{i+1}_{I_j})],
		\end{equation*}
		{where, for simplicity of notation, we have dropped the dependence on $n$ of the angles $\theta^i_{I_j}$. Taking into account the previous formulae,} we gain
		{
			\begin{align}
				\label{eqqnn1}
				& \mathcal{G}_n(h_n) =\lambda_n\sum_{j=1}^M\sum_{i\in\mathcal{I}^n(I_j)}\left\{\alpha_n\left[R^2-\prodscal{\pi\widetilde{u}^i_{nI_j}}{\pi\widetilde{u}^{i+1}_{nI_j}}\right]-\left[R^2-\prodscal{\pi\widetilde{u}^{i}_{nI_j}}{\pi\widetilde{u}^{i+2}_{nI_j}}\right]\right\}\notag\\
				& + \lambda_n R^2\left(1+\frac{\alpha_n^2}{8}\right)(\#\mathcal{I}^n(I)-M+1)+\lambda_n R^2(1-\alpha_n)\sum_{j=1}^M\#\mathcal{I}^n(I_j)\notag\\
				& =\lambda_nR^2\sum_{j=1}^M\sum_{i\in\mathcal{I}^n(I_j)}\left\{2\alpha_n\sin^2\bigg( \frac{\theta^i_{I_j}}{2} \bigg)-\big[1-\cos(\theta^i_{I_j}+\theta^{i+1}_{I_j})\big]\right\}\notag\\
				& + \lambda_n R^2\left(2-\alpha_n+\frac{\alpha_n^2}{8}\right)(\#\mathcal{I}^n(I)-M+1).
			\end{align}
		}
		{
			The proof can be carried out as in \cite[Theorem 4.2]{CicaleseSolombrino2015}. For reader’s convenience we give here its sketch. By trigonometric identities, it holds
			\begin{equation*}
				8\sin^2\bigg( \frac{\theta^i_{I_j}}{2} \bigg)-2\sin^2(\theta^i_{I_j})=8\sin^4\bigg( \frac{\theta^i_{I_j}}{2} \bigg).
			\end{equation*}
			Moreover, taking into account the boundary condition \eqref{BDcond}, we can find a vanishing sequence $\{\gamma_n\}_{n\in\N}\subset\R$ such that
			\begin{align*}
				& \sum_{i\in\mathcal{I}^n(I_j)}\big[2\sin^2(\theta^i_{I_j})-1+\cos(\theta^i_{I_j}+\theta^{i+1}_{I_j})\big]\\
				& \geq 2(1-\gamma_n)\sum_{i\in\mathcal{I}^n(I_j)}\bigg(\sin\bigg( \frac{\theta^{i+1}_{I_j}}{2} \bigg)-\sin\bigg( \frac{\theta^i_{I_j}}{2} \bigg)\bigg)^2.
			\end{align*}
			We insert the previous two formulae in \eqref{eqqnn1} and compute
			\begin{align*}
				& \mathcal{G}_n(h_n)\\
				&  =\lambda_n R^2 \sum_{j=1}^M\sum_{i\in\mathcal{I}^n(I_j)}\Bigg\{8\sin^2\bigg( \frac{\theta^i_{I_j}}{2} \bigg)-(8-2\alpha_n)\sin^2\bigg( \frac{\theta^i_{I_j}}{2} \bigg)+2\bigg(1-\frac{\alpha_n}{4}\bigg)^2\Bigg\}\\
				& -\lambda_n R^2 \sum_{j=1}^M\sum_{i\in\mathcal{I}^n(I_j)}\big[1-\cos(\theta^i_{I_j}+\theta^{i+1}_{I_j})\big]\\
				& -2\lambda_n R^2\bigg(1-\frac{\alpha_n}{4}\bigg)^2(\#\mathcal{I}^n(I)-M+1) + \lambda_n R^2\left(2-\alpha_n+\frac{\alpha_n^2}{8}\right)(\#\mathcal{I}^n(I)-M+1)\\
				& = \lambda_n R^2 \sum_{j=1}^M\sum_{i\in\mathcal{I}^n(I_j)}\Bigg\{8\sin^2\bigg( \frac{\theta^i_{I_j}}{2} \bigg)-2\sin^2(\theta^i_{I_j})-(8-2\alpha_n)\sin^2\bigg( \frac{\theta^i_{I_j}}{2} \bigg)+2\bigg(1-\frac{\alpha_n}{4}\bigg)^2\Bigg\}\\
				& +\lambda_n R^2 \sum_{j=1}^M\sum_{i\in\mathcal{I}^n(I_j)}\big[2\sin^2(\theta^i_{I_j})-1+\cos(\theta^i_{I_j}+\theta^{i+1}_{I_j})\big]\\
				& =8\lambda_n R^2\sum_{j=1}^M\sum_{i\in\mathcal{I}^n(I_j)}\bigg[\sin^2\bigg( \frac{\theta^i_{I_j}}{2} \bigg)-\frac{1}{2}\bigg(1-\frac{\alpha_n}{4}\bigg)\bigg]^2\\
				& +\lambda_n R^2 \sum_{j=1}^M\sum_{i\in\mathcal{I}^n(I_j)}\big[2\sin^2(\theta^i_{I_j})-1+\cos(\theta^i_{I_j}+\theta^{i+1}_{I_j})\big]\\
				& \geq \lambda_n R^2\sum_{j=1}^M\sum_{i\in\mathcal{I}^n(I_j)}\Bigg\{8\bigg[\sin^2\bigg( \frac{\theta^i_{I_j}}{2} \bigg)-\frac{1}{2}\bigg(1-\frac{\alpha_n}{4}\bigg)\bigg]^2+2(1-\gamma_n)\bigg[\sin\bigg( \frac{\theta^{i+1}_{I_j}}{2} \bigg)-\sin\bigg( \frac{\theta^i_{I_j}}{2} \bigg)\bigg]^2\Bigg\}.
			\end{align*}
			Dividing by $\sqrt{2}\lambda_n\delta_n^\frac{3}{2}$ and recalling that $\alpha_n=4(1-\delta_n)$, we infer that
			\begin{equation}
				\label{eqq1}
				\frac{\mathcal{G}_n(h_n)}{\sqrt{2}\lambda_n\delta_n^{\frac{3}{2}}}\geq R^2\Bigg\{ \frac{\sqrt{2}\delta_n^{\frac{1}{2}}}{\lambda_n}\sum_{i\in\mathcal{I}^n(I_j)}\lambda_n\big[(w_{nI_j}^i)^2-1\big]^2+\frac{\lambda_n}{\sqrt{2}\delta_n^{\frac{1}{2}}}(1-\gamma_n)\sum_{i\in\mathcal{I}^n(I_j)}\bigg(\frac{w_{nI_j}^{i+1}-w_{nI_j}^i}{\lambda_n}\bigg)^2\Bigg\}.
			\end{equation}
		}
		
		If $\varepsilon>0$ is sufficiently small {such} that $I_j^\varepsilon:=({t_j}+\varepsilon, {t_{j+1}}-\varepsilon) \subset (I_j)_n$, for all $n\in\N$, then
		\begin{equation*}
			MM_{n}(w_{n| I_j^\varepsilon}) \leq \lambda_n \delta_n^{\frac{3}{2}} C
		\end{equation*}
		and \eqref{eqq1} holds with $I_j^\varepsilon$ in place of $I_j$, {for any $j\in\{1,\dots,M\}$.} Therefore, {applying \cite[Theorem 2.2 and Remark 2.3]{CicaleseSolombrino2015} (see also \cite{Bra-Yip})
			,} $\{ w_{n} \chi_{I_j^\varepsilon} \}_{n \in \N} $ {converges}, up to subsequences, to $w \in BV(I_j)$ in $\mathrm{L}^1$. Thus we deduce the existence of $ h \in BV(I; \{-1,1\}\times\{ v_1,v_2\})	$ such that $h_n:=(w_n, \mathcal{A}(u_n)) \Ltheta h$.\\
		\indent Now we prove (ii). Let $h=(w,\mathcal{A}(u))\in BV(I;\{-1,1\}\times\{v_1,v_2\})$ {and $ \{ h_n\}_{n \in \N}\subset \widetilde{PC}_{\lambda_n}$ be such that $ h_n \Ltheta h$
			and \eqref{compEnglimitataordi12} holds true
			for some constant $C >0$. } {Up} to a subsequence, $M=M(h_n)$ is independent of $n$. {Moreover, denoting $I_j=(t_j, t_{j+1})$,} for $\varepsilon>0$ sufficiently small, {it holds that} $I_j^\varepsilon:=(t_j+ \varepsilon, t_{j+1}-\varepsilon) \subset (I_j)_n$, for all $ j\in\{1,\dots,M(h_n)\}$ and $n\in\N$. By {the definition of $\mathcal{G}_n$,} we have 
		\begin{equation*}
			\liminf_{n \rightarrow + \infty} \frac{{\mathcal{G}_n}(h_n)}{\sqrt{2}\lambda_n\delta_n^{\frac{3}{2}}}= \liminf_{n \rightarrow + \infty} \frac{\displaystyle\sum_{j=1}^{M} MM_n(h_{n|I_j})}{\sqrt{2}\lambda_n\delta_n^{\frac{3}{2}}}  \geq {\frac{4}{3}R^2} \sum_{j=1}^{M}\left| Dw \right|(I_j^\varepsilon ),
		\end{equation*}
		where in the last step we have used the liminf inequality of \cite[Theorem 4.2]{CicaleseSolombrino2015}. Letting $\varepsilon\rightarrow0$, we obtain the liminf inequality. \\
		\indent We finally prove (iii).	Let $h=(w,\mathcal{A}(u)) \in BV(I;\{-1,1\}\times\{ v_1,\,v_2\})$. We can find $M>0$ and an open partition of $I$ made by the intervals $ \mathcal{C}=\{I_j\}_{j\in\{1,\dots,M\}}$ such that $ h_{|I_j} =({w_{|I_j}, \overline{v}_j)\in BV(I_j;\{-1,1\})\times\{v_1,\, v_2\})}$. Thanks to the limsup inequality proved in \cite[Theorem 4.2]{CicaleseSolombrino2015}, for all $j \in \{1, \dots, M\}$ there exists a sequence $\{(z_j)_n\}_{n \in \N}\subset\mathrm{L}^1(I_j;\R)$, such that $(z_j)_n\rightarrow {w_{|I_j}}$ in $ \mathrm{L}^1(I_j; \mathbb{R})$ and 
		\begin{equation}\label{ultimaformula1111}
			\quad \lim_{n \rightarrow + \infty} \frac{MM_n(h_{n|I_j})}{\sqrt{2} \lambda_n \delta_n^{\frac{3}{2}}}= {\frac{4}{3}R^2}  \vert D{w}| (I_j),
		\end{equation}
		where $h_{n|I_j}:=((z_j)_n, \overline{v}_j)$. {By the definition of $\mathcal{G}_n$} and \eqref{ultimaformula1111}  we gain 
		\begin{equation*}
			\lim_{n \rightarrow + \infty} \frac{{\mathcal{G}_n}(h_n)}{\sqrt{2}\lambda_n\delta_n^{\frac{3}{2}}}= \lim_{n \rightarrow + \infty} \frac{\displaystyle\sum_{j=1}^{M} MM_n(h_{n_{|I_j}})}{\sqrt{2}\lambda_n\delta_n^{\frac{3}{2}}}  = {\frac{4}{3}R^2}\sum_{j=1}^{M}\left| D{w} \right|(I_j),
		\end{equation*}
		that is the thesis.
	\end{proof}
	
	\section{Analysis of the two-dimensional model}
	\label{Analysis of the two-dimensional model}
	{In this section} we analyze the problem in the two-dimensional case. Therefore we need to introduce proper notation and new definitions.
	\subsection{Further notation and definitions}
	Let $\{\lambda_n\}_{n\in\N}\subset\R^+$ be a vanishing sequence of positive lattice spacings. Given $i,j\in\Z$, we denote {by} $Q_{\lambda_n}(i,j):=(\lambda_n i,\lambda_n j)+[0,\lambda_n)^2$ the half-open square with left-bottom corner in $(\lambda_n i,\lambda_n j)$. For a given set $S$, we introduce the class of {spin fields} with values in $S$ which are piecewise constant on the squares of the lattice $\lambda_n\Z^2$:
	\begin{equation*}
		\mathcal{PC}_{\lambda_n}(\R^2;S):=\{ {u}\colon\R^2\rightarrow S\,:\, {u}(x)={u}(\lambda_n i,\lambda_n j) \text{ for }x\in Q_{\lambda_n}(i,j)\}.
	\end{equation*}
	We will identify a function ${u}\in \mathcal{PC}_{\lambda_n}(\R^2;S)$ with the function defined on the points of the lattice $\lambda_n\Z^2$ given by $(i,j)\mapsto {u}^{i,j}:={u}(\lambda_n i,\lambda_n j)$, for $i,j\in\Z$. Conversely, given values ${u}^{i,j}\in S$ for $i,j\in\Z$, we define ${u}\in \mathcal{PC}_{\lambda_n}(\R^2;S)$ by {setting} ${u}(x):={u}^{i,j}$, for $x\in Q_{\lambda_n}(i,j)$.\\
	\indent {Furthermore, we define the projection function $\mathcal{A}\colon \mathcal{PC}_{\lambda_n}(\R^2;S_1\cup S_2)\rightarrow\mathrm{L}^\infty(\R^2;\{v_1,v_2\})$ by setting}
	\begin{equation*}
		{
			\mathcal{A}(u)(x)=
			\begin{cases}
				v_1\quad&\text{if }u(x)\in S_1,\\
				v_2 &\text{if }u(x)\in S_2,
			\end{cases}
			\quad \forall x\in \R^2.}
	\end{equation*}
	
	\indent  In this paper we will make use of the notion of $BVG$ regularity. $BVG$ domains and $BVG$ functions have been introduced in \cite{Pol07} (see also \cite[Section 3]{CicaleseFosterOrlando2019}).
	{ 
		\begin{definition}
			\label{BVGReg}
			Let $I\subset\R$ be an open set. We define the space of $BVG$ functions by
			\begin{equation*}
				BVG(I):=\{\phi\in W^{1,\infty}(I)\,:\,\D\phi\in BV(I)\}.
			\end{equation*}
			A bounded connected open set $\Omega\subset\R^2$ is called a $BVG$ domain if $\Omega$ can be described locally at its boundary as the epigraph of a $BVG$ function with respect to a suitable choice of the axes, i.e., if for every $x\in\dd\Omega$ there exist a neighborhood $U_x\subset\R^2$, a function $\psi_x\in BVG(\R)$ and an isometry $R_x\colon\R^2\rightarrow\R^2$ satisfying
			\begin{equation*}
				R_x(\Omega\cap U_x)=\{(y_1,y_2)\in\R^2\,:\,y_1>\psi_x(y_2)\}\cap R_x(U_x).
			\end{equation*}
		\end{definition}
	}
We remark that smooth domains and polygons are $BVG$ domains and $BVG$ domains are Lipschitz domains.\\
	\indent As in the one-dimensional case we observe that, if $u\in\mathcal{PC}_{\lambda_n}(\R^2;S_1\cup S_2)$, {then a bounded connected open} set $\Omega\subset\R^2$ can be uniquely partitioned in regions where the {spin field} $u$ takes {values} only in one of the two {circles}. {In other words,} there exist $M(u)\in\N$ and a collection of connected open sets, $\{C_s\}_{s\in\{1,\dots,M(u)\}}$, such that
	\begin{equation}
		\label{Partizionehp1}
		\{C_s\}_{s\in\{1,\dots,M(u)\}}\text{ is an open partition of }\Omega,
	\end{equation}
	\begin{equation}
		\label{Partizionehp2}
		{\text{either } u(C_s)\subset S_1\text{ or } u(C_s)\subset S_2 ,\text{ for any $s\in\{1,\dots,M(u)$\}}},
	\end{equation}
	{
		\begin{align}
			\label{Partizionehp3}
			& \text{if }u(C_{s_1})\times u(C_{s_2})\subset S_\ell\times S_\ell,\text{ for some $s_1,s_2\in\{1,\dots,M(u)\}$ and $\ell\in\{1,2\}$,}\\
			& \text{then }\overline{C}_{s_1}\cap \overline{C}_{s_2}\text{ has at most a finite number of points.}
		\end{align}
	}
	
	\noindent{The last two} properties imply that this partition is unique. {We remark that the sets $C_s$ are squares or union of squares. In particular, {\eqref{Partizionehp3}} ensures that $u$ maps two confining sets of the open partition in different circles, if their intersection contain edges of squares.}

	The following definition will be useful throughout the section.
	\begin{definition}
		Let $u\in\mathcal{PC}_{\lambda_n}(\R^2;S_1\cup S_2)$ {and $\Omega\subset \R^2$ be a bounded connected open set}. We say that $\mathcal{C}_n(u)=\{ C_s\,|s\in\{1,\dots,M(u)\} \}$ is the open partition {of $\Omega$} associated with $u$ if  $M(u)\in\N$ and the collection $\{C_s\}_{s\in\{1,\dots,M(u)\}}$ of open connected sets satisfies \eqref{Partizionehp1}, \eqref{Partizionehp2} and \eqref{Partizionehp3}. {If $\Omega$ is a $BVG$ domain}, we call $\mathcal{C}_n(u)$ the open $BVG$ partition {of $\Omega$} associated with $u$ if $C_s$ is also {a} $BVG$ {domain}, for all $s\in\{1,\dots,M(u)\}$.
	\end{definition}
	\subsection{The energy model}
	Our model is an energy on discrete spin fields defined on square lattices inside a given domain $\Omega\subset\R^2$ belonging to the following class:
	\begin{equation*}
		\mathfrak{A}_0:=\{\Omega\subset\R^2\,:\,\Omega \text{ is a {simply connected $BVG$ domain}}\}{.}
	\end{equation*}
	To define the energies in our model, we introduce the set of indices
	\begin{equation*}
		\mathcal{I}^n(\Omega):=\{(i,j)\in\Z^2\,:\,\overline{Q}_{\lambda_n}(i,j), \overline{Q}_{\lambda_n}(i+1,j), \overline{Q}_{\lambda_n}(i,j+1)\subset\Omega \},
	\end{equation*}
	for $\Omega\in	\mathfrak{A}_0$. Let $\alpha_n:=4(1-\delta_n)$, where $\{\delta_n\}\subset\R^+$ is a vanishing sequence, and let $\{ k_n\}_{n\in\N}\subset\R^+$ {be} a divergent sequence. In the following we shall assume that $\varepsilon_n:=\frac{\lambda_n}{\sqrt{\delta_n}}\rightarrow0$ and $\lambda_n k_n\rightarrow\eta\in(0,+\infty)$, as $n\rightarrow +\infty$.\\
	\indent We consider the functionals $H_n,P_n\colon \mathrm{L}^\infty(\R^2;S_1\cup S_2)\times	\mathfrak{A}_0\rightarrow[0,+\infty]$ defined by
	\begin{equation*}
		H_n(u;\Omega):=\frac{1}{\sqrt{2}\lambda_n\delta_n^{\frac{3}{2}}}\frac{1}{2}\lambda_n^2\sum_{(i,j)\in\mathcal{I}^n(\Omega)}\left[\left| u^{i+2,j}-\frac{\alpha_n}{2}u^{i+1,j}+u^{i,j}\right|^2+\left| u^{i,j+2}-\frac{\alpha_n}{2}u^{i,j+1}+u^{i,j}\right|^2\right],
	\end{equation*}
	\begin{equation}
		P_n(u;\Omega):=\lambda_nk_n|D\mathcal{A}(u)|(\Omega),
	\end{equation}
	for $u\in\mathcal{PC}_{\lambda_n}(\R^2;S_1\cup S_2)$ and extended to $+\infty$ elsewhere.\\

	{Similarly to the analysis at the first and second order in the one-dimensional case,} we split the functional $H_n$ {as follows:}
	\begin{equation*}	H_n(u;\Omega)=\sum_{s=1}^{M(u)}\Big[\mathcal{H}_n(u;C_s)+(R_n)_{C_s}(u)\Big],
	\end{equation*}
	where 
	{
		\begin{equation*}
			\mathcal{H}_n(u;C_s):=H_n(u;C_s)+\frac{1}{\sqrt{2}\lambda_n\delta_n^{\frac{3}{2}}}\cdot 2\lambda_n^2(\alpha_n-1)(1-R^2)\#\mathcal{I}^n(C_s),
		\end{equation*}
	}
	\begin{align*}
		(R_n)_{C_s}(u):=
		& \frac{1}{\sqrt{2}\lambda_n\delta_n^{\frac{3}{2}}}\frac{1}{2}\lambda_n^2\sum_{(i,j)\in(C_s\cap\mathcal{I}^n(\Omega))\setminus\mathcal{I}^n(C_s)}\bigg[\left| u^{i+2,j}-\frac{\alpha_n}{2}u^{i+1,j}+u^{i,j}\right|^2\\
		&+\left| u^{i,j+2}-\frac{\alpha_n}{2}u^{i,j+1}+u^{i,j}\right|^2\bigg]
		{-\frac{1}{\sqrt{2}\lambda_n\delta_n^{\frac{3}{2}}}\cdot 2\lambda_n^2(\alpha_n-1)(1-R^2)\#\mathcal{I}^n(C_s),}
	\end{align*}
	{for any $s\in\{1,\dots,M(u)\}$.}
	{The functionals $(R_n)_{C_s}$ collect the remainders associated with the decomposition of the energy in the open partition $\mathcal{C}_n(u)=\{ C_s\,|s\in\{1,\dots,M(u)\} \}$. They consist of the interactions between spin field's vectors located in different circles.}\\
	\subsection{The $\Gamma$-convergence result}
	{In this subsection} we introduce the chirality order parameter associated with a spin field. Let $u\in\mathcal{PC}_{\lambda_n}(\R^2;S_1\cup S_2)$ and let $\mathcal{C}_n(u)=\{ C_s\,|s\in\{1,\dots,M(u)\} \}$ be the partition associated with $u$. For $(i,j)\in\mathcal{I}^n(C_s)$, {we consider the pairs $(u^{i,j},u^{i+1,j})$ and $(u^{i,j},u^{i,j+1})$ of vectors that take values in $S_\ell$, for some $\ell=\ell_s\in\{1,2\}$}.  {We} define the horizontal and vertical oriented angles between two adjacent spin vectors by
	\begin{equation*}
		\widetilde{\theta}^{i,j}_{C_s}:=\chi [\pi_{v_\ell^\bot}(u^{i,j}),  \pi_{v_\ell^\bot}(u^{i+1,j})] \arccos(\pi_{v_\ell^\bot}(u^{i,j})\cdot \pi_{v_\ell^\bot}(u^{i+1,j}))\in[-\pi,\pi),
	\end{equation*}
	\begin{equation*}
		\check{\theta}^{i,j}_{C_s}:=\chi [\pi_{v_\ell^\bot}(u^{i,j}),  \pi_{v_\ell^\bot}(u^{i,j+1})]\arccos(\pi_{v_\ell^\bot}(u^{i,j})\cdot \pi_{v_\ell^\bot}(u^{i,j+1}))\in [-\pi,\pi).
	\end{equation*}
	We define the order parameter $({(w,z)},\mathcal{A}(u))\in\mathcal{PC}_{\lambda_n}(\R^2;\R^2)\times{\mathrm{L}^\infty(\Omega;\{v_1,v_2\})}$ {(we will write $(w,z,\mathcal{A}(u))$ for simplicity)} {by setting}
	\begin{equation*}
		w^{i,j}:=
		\begin{cases}
			\sqrt{\frac{2}{\delta_n}}\sin\frac{\widetilde{\theta}^{i,j}_{C_s}}{2}\quad&\text{if }(i,j)\in\mathcal{I}^n(C_s)\text{ for some }s\in\{1,\dots,M(u)\},\\
			0 & \text{otherwise},
		\end{cases} 
	\end{equation*}
	\begin{equation*}
		z^{i,j}:=
		\begin{cases}
			\sqrt{\frac{2}{\delta_n}}\sin\frac{\check{\theta}^{i,j}_{C_s}}{2}\quad&\text{if }(i,j)\in\mathcal{I}^n(C_s)\text{ for some }s\in\{1,\dots,M(u)\},\\
			0 & \text{otherwise}.
		\end{cases} 
	\end{equation*}
	It is convenient to introduce the transformation $T_n\colon\mathcal{PC}_{\lambda_n}(\R^2;S_1\cup S_2)\rightarrow\mathcal{PC}_{\lambda_n}(\R^2;\R^2)\times\mathrm{L}^\infty(\Omega;\{v_1,v_2\})$ given by
	\begin{equation*}
		T_n(u):=\big(w,z,\mathcal{A}(u)).
	\end{equation*}
	\indent With a slight abuse of notation we define the functional $H_n\colon\mathrm{L}^1_{loc}(\R^2;\R^2\times\{v_1,v_2\})\times	\mathfrak{A}_0\rightarrow[0,+\infty)$ by {setting}
	\begin{equation}\label{31052023pom1}
		H_n(h;\Omega)=
		\begin{cases}
			H_n(u;\Omega) \quad&\text{if }T_n(u)=h\text{ for some }u\in\mathcal{PC}_{\lambda_n}(\R^2;S_1\cup S_2),\\
			+\infty&\text{otherwise}.
		\end{cases}
	\end{equation}
	Notice that $H_n$ does not depend on the particular choice of $u$, since it is rotation-invariant. The same notation can be adopted for $P_n$, $(R_n)_{C_s}$ and {$\mathcal{H}_n$.}\\
	\indent We study the convergence of the functional
	{
		\begin{align*}
			G_n(h;\Omega)
			& :=
			\begin{cases}
				\displaystyle  H_n(h,\Omega)-\sum_{s=1}^{M(h)} (R_n)_{C_s}(h)\quad&\text{if }T_n(u)=h\text{ for some }u\in\mathcal{PC}_{\lambda_n}(\R^2;S_1\cup S_2),\\
				+\infty &\text{otherwise}
			\end{cases}\\
			& =\sum_{s=1}^{M(h)}\mathcal{H}_n(h;C_s).
		\end{align*}
	}
	where $M(h):=M(u)$. Hence, we introduce the functional $G\colon\mathrm{L}^1_{loc}(\R^2;\R^2\times\{v_1,v_2\})\times	\mathfrak{A}_0\rightarrow [0,+\infty)$ by setting
	\begin{equation*}
		{G}(h;\Omega):=
		\begin{cases}
			\displaystyle\frac{4}{3}{R^2}\sum_{s=1}^{M(h)}(|D_1 w|(C_s)+|D_2 z|(C_s))\quad &\text{if }h=(w,z,\alpha)\in\text{Dom}({G};\Omega),\\
			+\infty &\text{otherwise},
		\end{cases}
	\end{equation*}
	where
	\begin{align*}
		\text{Dom}({G};\Omega):=
		& \bigg\{(w,z,\alpha)\in\mathrm{L}^1_{loc}(\R^2;\R^2\times\{v_1,v_2\}) \,:\,\exists\{C_s\}_{\substack{s\in\{1,\dots,M\}}}\text{ open partition of }\Omega\text{ s.t. }\\
		& (w_{|C_s},z_{|C_s},\alpha_{|C_s})\in BV(C_s;\{-1,1\}^2\times\{v_{\ell_s}\}),\,\text{for some } \ell_s\in\{1,2\},\\
		& \textnormal{curl}(w_{|C_s},z_{|C_s})=0\text{ in } \mathcal{D}'(C_s;\R^2) \bigg\}.
	\end{align*}
For $h\in\mathrm{Dom}(G;\Omega)$ we say that the collection $\{C_s\}_{s\in\{1,\dots,M\}}$ existing in virtue of the definition of $\mathrm{Dom}(G;\Omega)$ is the open partition associated with $h$.\\
\noindent We have denoted by $\mathcal{D}'(C_s;\R^2)$ the space of distributions and {by $\mathrm{curl}$ the distribution curl defined by}
	\begin{equation*}
		\langle (\textnormal{curl}(T))_{h,k}, \xi \rangle:=- \langle T^k, \partial_h \xi \rangle+ \langle T^h, \partial_k \xi \rangle, \quad {\forall} \xi \in C_c^{\infty}(C_s),\,\forall T \in \mathcal{D}'(C_s;\R^2),
	\end{equation*}
	for any $h,k\in\{1,2\}$.\\
	\indent {The following notion of convergence will be used.}
	\begin{definition}
		Let $\{h_n\}_{n\in\N}\subset \mathrm{L}^1_{loc}(\R^2;\R^2\times\{v_1,v_2\})$. We say that $h_n$ ${\Theta}$-converges to $h\in\mathrm{L}^1_{loc}(\R^2;\R^2\times\{v_1,v_2\})$ (we write $h_n\LTheta h$) if the following conditions are satisfied:
		\begin{itemize}
			\item there exist $\{u_n\}_{n\in\N}\subset\mathcal{PC}_{\lambda_n}(\R^2;S_1\cup S_2)$, a positive constant $C$ such that
			\begin{itemize}
				\item $h_n=T_n(u_n)$ and $P_n(u_n;\Omega)<C$,
				\item $M(u_n)\rightarrow M\in\N$ as $n\rightarrow+\infty$,
				\item $(C_s)_n\rightarrow C_s$ in the Hausdorff sense, as $n\rightarrow +\infty$, for any $s\in\{1,\dots,M\}$,
			\end{itemize}
			\noindent  where $\mathcal{C}_n(u_n)=\{(C_{s})_n|\,s\in\{1,\dots,M(h_n)\}\}$ is the open partition associated with $u_n$.
			\item ${h_n}\chi_{(C_s)_n}\rightarrow h\chi_{C_s}$ in $\mathrm{L}^1_{loc}(\R^2;\R^2\times\{v_1,v_2\})$, for any $s\in\{1,\dots,M\}$.
		\end{itemize}
	\end{definition}
	\noindent As in formula \eqref{31052023pom1} we define $ P_n(h;\Omega):=P_n(u;\Omega)$ for $ h= T_n(u)$ with $ u \in \mathcal{PC}_{\lambda_n}(\R^2;S_1\cup S_2)$.\\
	\indent We remark that in general it is not possible to prove a compactness result for a sequence $\{h_n=T_n(u_n)\}_{n\in\N}\subset T_n(\mathcal{PC}_{\lambda_n}(\R^2;S_1\cup S_2))$ satisfying only the following natural conditions:
	 \begin{equation*}
		{\sup_{n\in\N}G_n(h_n;\Omega)< C\quad\text{and}\quad \sup_{n\in\N}P_n(h_n;\Omega)<C}.
	\end{equation*}
	Indeed, it could happen that the region $\{\mathcal{A}(u_n)=v_1\}$ has an increasing number of holes vanishing in the limit so that $\{M(u_n)\}_{n\in\N}$ is divergent. Neither the Hausdorff convergence of the sets of the open partition is ensured.\\
	\indent In the following proposition we show that, if strong and technical conditions hold, then $\{h_n\}_{n\in\N}$ converges, up to subsequences, with respect to the $\Theta$-convergence.
    
	\begin{proposition}
		Let $\{h_n=T_n(u_n)\}_{n\in\N}\subset T_n(\mathcal{PC}_{\lambda_n}(\R^2;S_1\cup S_2))$ be a sequence such that
		\begin{equation}
			\label{comp}
			{\sup_{n\in\N}G_n(h_n;\Omega)< C\quad\text{and}\quad \sup_{n\in\N}P_n(h_n;\Omega)<C},
		\end{equation}
		for some constant $C>0$. Furthermore, we assume that the open partition associated with $u_n$, $\mathcal{C}_n(u_n)=\{(C_{s})_n|\,s\in\{1,\dots,M(u_n)\}\}$, is such that
		\begin{equation*}
			M(u_n)\rightarrow M\in\N\quad\text{as }n\rightarrow+\infty,
		\end{equation*}
		\begin{equation*}
			(C_s)_n\rightarrow C_s \quad\text{in the Hausdorff sense, as }n\rightarrow +\infty,\,\forall s\in\{1,\dots,M\}.
		\end{equation*} Then there exists $h\in\textnormal{Dom}({G};\Omega)$ such that, up to a subsequence, $h_n\LTheta h$.
	\end{proposition}
	
\begin{proof}
	 Let $\{h_n=(w_n,z_n,\mathcal{A}(u_n))\}_{n\in\N}\subset T_n(\mathcal{PC}_{\lambda_n}(\R^2;S_1\cup S_2))$ be a sequence satisfying \eqref{comp}. {Since ${u_n}_{|C_s}\in S_\ell$, for some $\ell=\ell_s\in\{1,2\}$, then, 
		by geometric and trigonometric identities, we deduce that
		\begin{equation*}
			\prodscal{u^{i,j}}{u^{i+1,j}}=1-R^2+\prodscal{\pi u^{i,j}}{\pi u^{i+1,j}},
		\end{equation*}
		\begin{equation*}
			\prodscal{u^{i,j}}{u^{i,j+1}}=1-R^2+\prodscal{\pi u^{i,j}}{\pi u^{i,j+1}},
		\end{equation*}
		where $\pi u^{i,j}:=\pi_{v_\ell^\bot} u^i$. Thus we may write
		\begin{equation*}
			G_n(h_n;\Omega)=\sum_{s=1}^M \widetilde{H}_n(u_n;C_s),
		\end{equation*}
		where
		\begin{align*}
			&\widetilde{H}_n(h_n;C_s)\\
			&:=\frac{1}{\sqrt{2}\lambda_n\delta_n^{\frac{3}{2}}}\frac{1}{2}\lambda_n^2\sum_{(i,j)\in\mathcal{I}^n(C_s)}\left[\left| \pi u_n^{i+2,j}-\frac{\alpha_n}{2}\pi u_n^{i+1,j}+\pi u_n^{i,j}\right|^2+\left| \pi u_n^{i,j+2}-\frac{\alpha_n}{2}\pi u_n^{i,j+1}+\pi u_n^{i,j}\right|^2\right].
		\end{align*}	} Fixing $\varepsilon>0$ sufficiently small, we have that for all $n\in\N$, up to a subsequence, $(C_s)_\varepsilon:=\{x\in C_s\,:\,\text{dist}(x,\partial C_s)>\varepsilon\}\subset (C_s)_n$ and $u_{n_{|(C_s)_\varepsilon}}$ takes values only in one {circle}. We infer that
	\begin{equation*}
		{\sum_{s=1}^M \widetilde{H}_n(h_n;(C_s)_\varepsilon)}\leq {G}_n(h_n;\Omega)<C,
	\end{equation*}
	which of course implies that ${\widetilde{H}}_n(h_n;(C_s)_\varepsilon)<C$, for all $s\in\{1,\dots,M\}$. We are in position to apply \cite[Theorem 2.1 i) and Remark 2.2]{CicaleseFosterOrlando2019} to deduce the existence of $(w_{(C_s)_\varepsilon},z_{(C_s)_\varepsilon})\in BV((C_s)_\varepsilon;\{-1,1\}^2)$ such that, up to subsequences, $(w_n,z_n)\rightarrow(w_{(C_s)_\varepsilon},z_{(C_s)_\varepsilon})$ in $\mathrm{L}^1_{loc}((C_s)_\varepsilon;\R^2)$ and curl$(w_{(C_s)_\varepsilon},z_{(C_s)_\varepsilon})=0$ in $\mathcal{D}'((C_s)_\varepsilon;\R^2)$. The couples $(w_{(C_s)_\varepsilon},z_{(C_s)_\varepsilon})$ can be extended to 0 in $C_s\setminus(C_s)_\varepsilon$. We preliminary observe that
	\begin{equation}
		\label{eq11}
		(w_{(C_s)_{\varepsilon_2}},z_{(C_s)_{\varepsilon_2}})
		=(w_{(C_s)_{\varepsilon_1}},z_{(C_s)_{\varepsilon_1}}) \quad\text{a.e. on }(C_s)_{\varepsilon_2},
	\end{equation}
	for any $0<\varepsilon_1<\varepsilon_2$. Indeed, since $(C_s)_{\varepsilon_2}\subset(C_s)_{\varepsilon_1}$, we have that
	\begin{equation*}
		(w_n,z_n)\rightarrow (w_{(C_s)_{\varepsilon_1}},z_{(C_s)_{\varepsilon_1}})\quad\text{in}\mathrm{L}^1_{loc}((C_s)_{\varepsilon_2};\R^2).
	\end{equation*}
	The uniqueness of the limit in the $\mathrm{L}^1_{loc}$-topology implies \eqref{eq11}. We now define the couples $(w_{C_s},z_{C_s})\colon C_s\rightarrow\R^2$ by
	\begin{equation*}
		(w_{C_s},z_{C_s}):=\lim_{\varepsilon\rightarrow 0^+}(w_{(C_s)_\varepsilon},z_{(C_s)_\varepsilon}).
	\end{equation*}
	The definition is well-posed; indeed, since by \eqref{eq11},
	\begin{equation*}
		\lim_{\varepsilon'\rightarrow 0^+}(w_{(C_s)_{\varepsilon'}},z_{(C_s)_{\varepsilon'}})=(w_{(C_s)_{\frac{1}{n}}},z_{(C_s)_{\frac{1}{n}}}) \quad\text{a.e. in }(C_s)_{\frac{1}{n}},
	\end{equation*}
	for all $n\in\N$, then
	\begin{align*}
		&\left|\left\{ x\in C_s\,:\,\nexists\lim_{\varepsilon'\rightarrow 0^+}(w_{(C_s)_{\varepsilon'}}(x),z_{(C_s)_{\varepsilon'}}(x)) \right\}\right|\\
		&=\left|\bigcup_{n=1}^{+\infty}\left\{ x\in(C_s)_{\frac{1}{n}}\,:\,\nexists\lim_{\varepsilon'\rightarrow 0^+}(w_{(C_s)_{\varepsilon'}}(x),z_{(C_s)_{\varepsilon'}}(x))  \right\}\right|=0.
	\end{align*}
	Furthermore we {define} $(w,z)\colon\Omega\rightarrow\R^2$ {by setting}
	\begin{equation}
		(w,z)(x)=(w_{C_s},z_{C_s})(x),
	\end{equation}
	for a.e. $x\in\Omega$ with $x\in C_s$, for some $s\in\{1,\dots,M\}$. Of course $(w_{|C_s},z_{|C_s})=(w_{C_s},z_{C_s})\in BV(C_s;\{-1,1\}^2)$, as it is the limit of $BV$ functions. In order to show the $\mathrm{L}^1_{loc}$-convergence, we fix $A\subset\subset C_s$. Since dist$(A,\dd C_s)>0$, there exists $\varepsilon>0$ such that  $A\subset\subset (C_s)_\varepsilon$. We obtain:
	\begin{equation*}
		\norm{(w_n,z_n)-(w_{C_s},z_{C_s})}_{\mathrm{L}^1(A;\R^2)}=\norm{(w_n,z_n)-(w_{(C_s)_\varepsilon},z_{(C_s)_\varepsilon})}_{\mathrm{L}^1(A;\R^2)},
	\end{equation*}
	which vanishes as $n\rightarrow+\infty$, up to subsequences. This leads to the convergence
	\begin{equation*}
		(w_n,z_n)\rightarrow( w_{C_s},z_{C_s}) \quad\text{in }L^1_{loc}(C_s;\R^2).
	\end{equation*}
	Finally, we prove that $\mathrm{curl}(w_{C_s},z_{C_s})=0$ in $\mathcal{D}'(C_s;\R^2)$. If $\xi\in C_c^\infty(C_s)$, then $\mathrm{supp}\xi\subset(C_s)_\varepsilon$ for some $\varepsilon>0$ and so
	\begin{align*}
		\left\langle \textnormal{curl}(w_{C_s},z_{C_s}),\xi \right\rangle
		&=-\int_{(C_s)_\varepsilon}w_{(C_s)_\varepsilon}\dd_2\xi\,dx+\int_{(C_s)_\varepsilon}z_{(C_s)_\varepsilon}\dd_1\xi\,dx\\
		&=\left\langle \textnormal{curl}(w_{(C_s)_\varepsilon},z_{(C_s)_\varepsilon}),\xi \right\rangle=0.
	\end{align*}
\end{proof}

	Now we state the main theorem of this section. The regularity assumption on $\Omega$ and on the open partition of $h$ in the statement ii) are required in order to apply \cite[Theorem 2.1 iii)]{CicaleseFosterOrlando2019} locally. As explained in \cite{CicaleseFosterOrlando2019} a simply connected $BVG$ domain guaranties an extension property for $BVG$ functions, which is needed to construct a recovery sequence for $h$. On the contrary, the proof of the liminf inequality i) actually works without assuming this kind of regularity (see \cite[Remark 2.2]{CicaleseFosterOrlando2019}).
	\begin{theorem}
		\label{Teorema principale 2d}
		Let $\Omega\in	\mathfrak{A}_0$. Then the following statements hold true:
		\begin{itemize}
			\item[i)]\textit{(liminf inequality)} Let $\{h_n\}_{n\in\N}\subset\mathrm{L}^1_{loc}(\R^2;\R^2\times\{v_1,v_2\})$ and $h\in\mathrm{L}^1_{loc}(\R^2;\R^2\times\{v_1,v_2\})$. Assume that $\displaystyle{\sup_{n\in\N}}P_n(h_n;\Omega)<C$ for some constant $C>0$ and $h_n\LTheta h$. Then
			\begin{equation*}
				{G}(h;\Omega)\leq\liminf_{n\rightarrow+\infty}{G}_n(h_n;\Omega).
			\end{equation*}
			\item[ii)]\textit{(limsup inequality)} Let $h\in\mathrm{Dom}(G;\Omega)$ be such that its open partition consists of BVG domains. Then there exists a sequence $\{h_n\}_{n\in\N}\subset\mathrm{L}^1_{loc}(\R^2;\R^2\times\{v_1,v_2\})$ such that $h_n\LTheta h$ and
			\begin{equation*}
				\limsup_{n\rightarrow+\infty}{G}_n(h_n;\Omega)\leq{G}(h;\Omega).
			\end{equation*}
		\end{itemize}
	\end{theorem}
	\begin{proof}
		We start {by} proving i). 
 Let $\{h_n\}_{n\in\N}\subset\mathrm{L}^1_{loc}(\R^2;\R^2\times\{v_1,v_2\})$ and $h\in\mathrm{L}^1_{loc}(\R^2;\R^2\times\{v_1,v_2\})$ {be} such that $\displaystyle{\sup_{n\in\N}}P_n(h_n;\Omega)<C$ and $h_n\LTheta h$. Up to subsequences, we {may} assume that the lower limit in the right hand side of the liminf inequality is actually a limit. Furthermore we {may} assume that it is finite, the inequality being otherwise trivial. In particular, we have
		\begin{equation*}
			{\sup_{n\in\N}G_n}(h_n;\Omega)<C,
		\end{equation*}
		possibly with a larger $C$. By the definition of ${\Theta}$-convergence, $h_n=(w_n,z_n,\mathcal{A}(u_n))=T_n(u_n)$ for some $u_n\in\mathcal{PC}_{\lambda_n}(\R^2;S_1\cup S_2)$. Up to subsequences, $M=M(h_n)$ is independent of $n$ and {we may assume}, for $\varepsilon>0$ sufficiently small, that $(C_s)_\varepsilon\subset (C_s)_n$  and $u_{n_{|(C_s)_\varepsilon}}$ takes values only on one {circle $S_\ell$}, for all $n\in\N$. Reasoning as in i), we infer
		\begin{equation*}
			{{G}_n(h_n;\Omega)\geq\sum_{s=1}^M \widetilde{H}_n(h_n;(C_s)_\varepsilon)},
		\end{equation*}
		Since ${h_n}\rightarrow h$ in $\mathrm{L}^1((C_s)_\varepsilon;\R^2\times\{v_\ell\})$, as $n\rightarrow+\infty$, we {are in position to} apply \cite[Theorem 2.1 ii) and Remark 2.2]{CicaleseFosterOrlando2019} so that, passing to the lower limit,  we get
		\begin{align*}
			\liminf_{n\rightarrow +\infty}{G}_n(h_n;\Omega)
			&\geq \sum_{s=1}^M\liminf_{n\rightarrow +\infty}{\widetilde{H}}_n(h_n;(C_s)_\varepsilon)
			\geq \sum_{s=1}^M \frac{4}{3}R^2[|D_1 w|((C_s)_\varepsilon)+|D_2 z|((C_s)_\varepsilon)],
		\end{align*}
		where $h=(w,z,\alpha)$. Letting $\varepsilon\rightarrow 0^+$ we get the thesis.\\
		\indent Let us prove ii). Let $h\in\text{Dom}(G;\Omega)$. This implies that $h=(w,z,\alpha)\in\mathrm{L}^1_{loc}(\R^2;\R^2\times\{v_1,v_2\})$ and the existence of an open partition of $\Omega$, $\mathcal{C}=\{C_{s}|\,s\in\{1,\dots,M\}\}$ consisting of $BVG$ domains such that, {for some $\ell=\ell_s\in\{1,2\}$},
		\begin{equation*}
			(w_{|C_s},z_{|C_s},\alpha_{|C_s})\in BV(C_s;\{-1,1\}^2\times\{v_\ell\})\quad\text{and}\quad\textnormal{curl}(w_{|C_s},z_{|C_s})=0\text{ in } \mathcal{D}'(C_s;\R^2).
		\end{equation*}
		Applying \cite[Theorem 2.1 iii)]{CicaleseFosterOrlando2019} to any $(w_{|C_s},z_{|C_s})$, we get the existence of a sequence\break ${\{((w_n)_{C_s},(z_n)_{C_s})\}_{n\in\N}\subset}\mathrm{L}^1_{loc}(\R^2;\R^2)$ such that ${((w_n)_{C_s},(z_n)_{C_s})}\rightarrow (w_{|C_s},z_{|C_s})$ in $\mathrm{L}^1(C_s;\R^2)$ and
		\begin{equation*}
			\limsup_{n\rightarrow+\infty}{\mathcal{H}}_n({(w_n)_{C_s},(z_n)_{C_s}},v_\ell)\leq \frac{4}{3}{R^2}(|D_1 w|(C_s)+|D_2 z|(C_s))
		\end{equation*}
		Defining $(w_n,z_n,\alpha_n)\colon\R^2\rightarrow\R^2\times\{v_1,v_2\}$ by
		\begin{equation*}
			(w_n,z_n,\alpha_n)(x):={((w_n)_{C_s}(x),(z_n)_{C_s}(x),v_\ell)},
		\end{equation*}
		if $x\in\Omega$ such that $x\in C_s$ for some $s\in\{1,\dots,M\}$, and arbitrarily extended outside $\Omega$, and summing on $s\in\{1,\dots,M\}$ the previous inequality we obtain the thesis.
	\end{proof}
	
\section*{Acknowledgments}
	The authors wish to thank the reviewer for numerous suggestions that improved the paper.
	The authors warmly thank Prof. Marco Cicalese for the insightful discussions.
	L. Lamberti wishes to acknowledge the hospitality of the Faculty of Mathematics of the Technical University of Munich, where part of this research was carried out. The authors are members of the Gruppo Nazionale per l’Analisi Matematica, la Probabilità e le loro Applicazioni (GNAMPA) of the Istituto Nazionale di Alta Matematica (INdAM).  
	A. Kubin was supported by the DFG Collaborative Research Center TRR 109 “Discretization in Geometry and Dynamics”. L. Lamberti was supported partially by the DFG Collaborative Research Center TRR 109 “Discretization in Geometry and Dynamics” and by COST Action 18232 MAT-DYN-NET, supported by COST (European Cooperation in Science and Technology).

\section*{Conflict of interest}
The authors  declare no conflict of interest.


\begin{thebibliography}{999}

		\bibitem{ACXY}
		R. Alicandro, M. Cicalese, Variational analysis of the asymptotics of the ${\MakeUppercase
			{xy}}$ model, {\em Arch. Rat. Mech. Anal.}, \textbf{192} (2009), 501--36. https://doi.org/10.1007/s00205-008-0146-0
		
		\bibitem{AlicandroCicalseGloria2008} R. Alicandro, M. Cicalese, A. Gloria, Variational description of bulk energies for bounded and unbounded spin systems, \emph{Nonlinearity}, \textbf{21} (2008), 1881--1910. http://dx.doi.org/10.1088/0951-7715/21/8/008
		
		\bibitem{AmbFusPal}
		L. Ambrosio, N. Fusco, D. Pallara,  Functions of bounded	variation and free discontinuity problems, Oxford Mathematical Monographs,
		The Clarendon Press, (2000).
		
		\bibitem{BCKO} A. Bach, M. Cicalese, L. Kreutz, G. Orlando, The antiferromagnetic $XY$ model on the triangular lattice: chirality transitions at the surface scaling, \emph{Calc. Var.}, \textbf{60} (2021), 149. https://doi.org/10.1007/s00526-021-02016-3
		
		\bibitem{BadCicDLPon} R. Badal, M. Cicalese, L. De Luca, M. Ponsiglione, $\Gamma$-convergence analysis of a generalized $XY$ model: fractional vortices and string defects, \emph{Comm. Math. Phys.}, \textbf{358} (2018), 705--739. https://doi.org/10.1007/s00220-017-3026-3
		
		\bibitem{GCB} A. Braides.
		$\Gamma$-convergence for beginners, Oxford University Press, (2002).
		
		\bibitem{BraTru}
		A. Braides, L. Truskinovsky, Asymptotic expansions by
		$\Gamma$-convergence, \emph{Contin. Mech. Thermodyn.}, \textbf{20} (2008), 21--62. https://doi.org/10.1007/s00161-008-0072-2
		
		\bibitem{Bra-Yip}
		A. Braides, N. K. Yip, A quantitative description of mesh	dependence for the discretization of singularly perturbed nonconvex problems, \emph{SIAM J. Numer. Anal.}, \textbf{50} (2012), 1883--1898. https://doi.org/10.1137/110822001
		
		\bibitem{CicaleseFosterOrlando2019}M. Cicalese, M. Forster, G. Orlando, Variational analysis of a two-dimensional frustrated spin system: emergence and rigidity of chirality transitions, \emph{SIAM J. Math. Anal.}, \textbf{51} (2019), 4848--4893. https://doi.org/10.1137/19M1257305
		
		\bibitem{CicOrlRuf}
		M. Cicalese, G. Orlando, M. Ruf, Emergence of concentration effects in the variational analysis of the $N$-clock model, \emph{Commun. Pure Appl. Anal.} \textbf{75} (2019), 2279--2342. https://doi.org/10.1002/cpa.22033
		
		\bibitem{CRS} M. Cicalese, M. Ruf, F. Solombrino, Chirality transitions in frustrated S$^2$-valued spin systems, \emph{Math. Models Methods Appl. Sci.}, \textbf{26} (2016), 1481--1529. https://doi.org/10.1142/S0218202516500366
		
		\bibitem{CicaleseSolombrino2015} M. Cicalese, F. Solombrino, Frustrated ferromagnetic spin chains: a variational approach to chirality transitions, \emph{J. Nonlinear Sci.}, \textbf{25} (2015), 291--313. https://doi.org/10.1007/s00332-015-9230-4
		
		
		\bibitem{diep}
		H. T. Diep, Frustrated spin systems, World Scientific, (2005).
		
		\bibitem{Dis} {R. S. Dissanayaka Mudiyanselage, H. Wang, O. Vilella , M. Mourigal, G. Kotliar et al., LiYbSe2: Frustrated Magnetism in the Pyrochlore Lattice, \emph{J. Am. Chem. Soc.}, \textbf{144}  (2022), 11933--11937.} https://pubs.acs.org/doi/10.1021/jacs.2c02839
		
		\bibitem{DmiKri}
		D. V. Dmitriev, V. Ya Krivnov, Universal low-temperature properties of frustrated classical spin chain near the ferromagnet-helimagnet transition point,  {\em EPJ B}, \textbf{82} (2011), 123--131.
		https://doi.org/10.1140/epjb/e2011-10664-6
		
		\bibitem{Dre-etal}
		S. L. Drechsler, O. Volkova, A.N. Vasiliev, N. Tristan, J. Richter, M. Schmitt et al., Frustrated cuprate route from antiferromagnetic to ferromagnetic spin-$\frac12$ Heisenberg chains: Li$_{2}$ZrCuO$_{4}$ as a missing link near the quantum critical point, {\em Phys. Rev. Lett.}, \textbf{98} (2007), 077202. https://doi.org/10.1103/PhysRevLett.98.077202
		
		
		\bibitem{GMc} {M. J. P. Gingras, P. A. McClarty, Quantum spin ice: a search for gapless quantum spin liquids in pyrochlore magnets, \emph{Rep. Prog. Phys.}, \textbf{77} (2014), 056501.} https://doi.org/10.1088/0034-4885/77/5/056501
		
		\bibitem{modica}
		L. Modica, The gradient theory of phase transitions and the minimal interface criterion, {\em Arch. Rat. Mech. Anal.}, \textbf{98} (1987), 123--142. https://doi.org/10.1007/BF00251230
		
		\bibitem{Mod-Mor}
		L. Modica, S. Mortola, Il limite nella $\Gamma$-convergenza di una famiglia di funzionali ellittici, {\em Boll. Un. Mat. Ital. A }, \textbf{14} (1977), 285--299.
		
		\bibitem{Noc} {D. G. Nocera, B. M. Bartlett,  D. Grohol, D. Papoutsakis, M. P. Shores, Spin frustration in 2D kagom{\'e} lattices: a problem for inorganic synthetic chemistry, \emph{Chem. Eur. J.}, \textbf{10} (2004), 3850--3859.} https://doi.org/10.1002/chem.200306074
		
		\bibitem{Pol07} A. Poliakovsky, Upper bounds for singular perturbation problems involving gradient fields, \emph{J. Eur. Math. Soc. (JEMS)}, \textbf{9} (2007), 1--43. http://dx.doi.org/10.4171/JEMS/70
		
		\bibitem{Pov} {K. Yu. Povarov, L. Facheris, S. Velja, D. Blosser, Z. Yan, S. Gvasaliya et al., Magnetization plateaux cascade in the frustrated quantum antiferromagnet Cs$_2$CoBr$_4$, \emph{Phys. Rev. Research}, \textbf{2} (2020), 043384.} https://doi.org/10.1103/PhysRevResearch.2.043384
		
		\bibitem{skomski2008} {R. Skomski, Simple models of magnetism. Oxford University Press on Demand, (2008).}
		
		\bibitem{SAAB} {R. Szymczak, P. Aleshkevych, C. P. Adams, S. N. Barilo, A. J. Berlinsky, J. P. Clancy et al., Magnetic anisotropy in geometrically frustrated kagomé staircase lattices, \emph{J. Magn. Magn. Mater.}, \textbf{321} (2009), 793.} https://doi.org/10.1016/j.jmmm.2008.11.076
		

\end{thebibliography}
\end{document}